\documentclass[acmsmall]{acmart}
\usepackage[utf8]{inputenc}
\usepackage{newunicodechar}
\usepackage{amsbsy}
\usepackage{pifont}
\usepackage{multirow}
\usepackage{listings}
\usepackage{xcolor}
\usepackage{parcolumns}
\usepackage[most]{tcolorbox}

\definecolor{eclipseStrings}{RGB}{42,0.0,255}
\definecolor{eclipseKeywords}{RGB}{127,0,85}
\colorlet{numb}{magenta!60!black}
\definecolor{background}{HTML}{EEEEEE}

\DeclareUnicodeCharacter{00A0}{ }

\lstdefinelanguage{json}{
    basicstyle=\tiny,
    commentstyle=\color{eclipseStrings}, 
    stringstyle=\color{eclipseKeywords}, 
    numbers=left,
    numberstyle=\scriptsize,
    stepnumber=1,
    numbersep=8pt,
    showstringspaces=false,
    breaklines=true,
    frame=lines,
    backgroundcolor=\color{background}, 
    string=[s]{"}{"},
    comment=[l]{:\ "},
    morecomment=[l]{:"},
    literate=
        *{0}{{{\color{numb}0}}}{1}
         {1}{{{\color{numb}1}}}{1}
         {2}{{{\color{numb}2}}}{1}
         {3}{{{\color{numb}3}}}{1}
         {4}{{{\color{numb}4}}}{1}
         {5}{{{\color{numb}5}}}{1}
         {6}{{{\color{numb}6}}}{1}
         {7}{{{\color{numb}7}}}{1}
         {8}{{{\color{numb}8}}}{1}
         {9}{{{\color{numb}9}}}{1}
}

\AtBeginDocument{%
  \providecommand\BibTeX{{%
    \normalfont B\kern-0.5em{\scshape i\kern-0.25em b}\kern-0.8em\TeX}}}

\copyrightyear{2022}
\setcopyright{acmcopyright}
\acmJournal{PACMHCI}
\acmYear{2022} \acmVolume{6} \acmNumber{EICS} \acmArticle{172} \acmMonth{6} \acmPrice{15.00}\acmDOI{10.1145/3534526}

\begin{document}

\title[Mod2Dash]{Mod2Dash: A Framework for Model-Driven Dashboards Generation}

\author{Liuyue Jiang}
\affiliation{%
  \institution{The Centre for Research on Engineering Software Technologies (CREST), University of Adelaide \& Cyber Security Cooperative Research Centre}
  \city{Adelaide}
  \country{Australia}}
\email{liuyue.jiang@adelaide.edu.au}

\author{Nguyen Khoi Tran}
\affiliation{%
  \institution{The Centre for Research on Engineering Software Technologies (CREST), University of Adelaide}
  \city{Adelaide}
  \country{Australia}}
\email{nguyen.tran@adelaide.edu.au}

\author{M. Ali Babar}
\affiliation{%
  \institution{The Centre for Research on Engineering Software Technologies (CREST), University of Adelaide \& Cyber Security Cooperative Research Centre}
  \city{Adelaide}
  \country{Australia}}
\email{ali.babar@adelaide.edu.au}

\renewcommand{\shortauthors}{Liuyue Jiang, Nguyen Khoi Tran, \& M. Ali Babar}


\begin{abstract}

The construction of an interactive dashboard involves deciding on what information to present and how to display it and implementing those design decisions to create an operational dashboard. Traditionally, a dashboard's design is implied in the deployed dashboard rather than captured explicitly as a digital artifact, preventing it from being backed up, version-controlled, and shared. Moreover, practitioners have to implement this implicit design manually by coding or configuring it on a dashboard platform. 
This paper proposes Mod2Dash, a software framework that enables practitioners to capture their dashboard designs as models and generate operational dashboards automatically from these models. The framework also provides a GUI-driven customization approach for practitioners to fine-tune the auto-generated dashboards and update their models. With these abilities, Mod2Dash enables practitioners to rapidly prototype and deploy dashboards for both operational and research purposes. We evaluated the framework's effectiveness in a case study on cyber security visualization for decision support. A proof-of-concept of Mod2Dash was employed to model and reconstruct 31 diverse real-world cyber security dashboards. A human-assisted comparison between the Mod2Dash-generated dashboards and the baseline dashboards shows a close matching, indicating the framework's effectiveness for real-world scenarios.

\end{abstract}



\maketitle

\section{Introduction}

\textit{"A dashboard is a visual display of the most important information needed to achieve one or more objectives that has been consolidated on a single computer screen so it can be monitored at a glance"} (Few, 2004 \cite{Few2004}).
Dashboards are considered effective means of presenting critical information by comprehensively using multiple visualization techniques to help decision makers facilitate decision making \cite{Rahman2017, Vazquez-Ingelmo2019slr, Sarikaya2019}. They are widely used in real-world environments across a variety of disciplines \cite{Rojas2020, DeCroon2021, Alzoubi2021}. 
For example, cyber security dashboards monitor and visualize cyber incidents, trends, and threats, to enhance cyber situational awareness (CSA) \cite{Jiang2021}.

Each dashboard encapsulates a set of design decisions for the specific practitioners.
The design decisions of what constitutes dashboards and how to visualize them implicitly reside in the mind of a dashboard practitioner as knowledge or in the deployed dashboards. A design decision is implied and not usually captured as a digital artifact, and the implicit design is not communicable.
So, collaboration with others to further improve the design and delivery to the development stage is challenging. Additionally, the implicit design cannot be version-controlled and backed up.

Traditionally, dashboard development involves phases such as users planning features and requirements, designers analyzing requirements and designing the visual interface, and developers coding and testing the dashboard \cite{Mohit2017}. As a result, capturing dashboard design and delivering a dashboard for the practitioners are time-consuming and challenging.

The dashboard platforms, such as Grafana\footnote{Grafana: https://grafana.com/} and Kibana\footnote{Kibana: https://www.elastic.co/kibana/}, allow dashboard practitioners to realize the dashboard design and create web-based dashboards without concerning about the implementation details.
However, these platforms are still manual in building the dashboard from individual components \cite{kibanatutorial, grafanatutorial}. The limited number of possible widgets, data sources, and features in these platforms also limit the imagination of dashboard design. 
Furthermore, they cannot capture the practitioners' implicit designs and automatically transform them into dashboards.

Therefore it is necessary to capture the dashboard design decisions.
The design decisions can be converted into a dashboard model, presenting what and how required information should be presented to enhance practitioners' situational awareness (SA) and decision making. 
Thus, dashboard development and delivery can be transformed into summarizing the design decisions as a model and implementing that model into an operational dashboard.

Using a dashboard model to capture design decisions, combined with the automated generation mechanism, can significantly reduce the work to create dashboards compared to building them from scratch \cite{Aksu2019}. Additionally, dashboard languages simplify communication and collaboration during the process of concretizing and implementing dashboard designs. Existing studies have also shown the usability of the dashboard model to represent dashboard designs in different disciplines (e.g., business performance \cite{Palpanas2007, Kintz2017} and knowledge management \cite{Vazquez-Ingelmo2020a}). However, their models fail to represent many crucial aspects of building real-world dashboards, such as dashboard layout, interactions, and customization features. Moreover, these dashboard languages are designed for different particular disciplines and have limited generalizability. Therefore, it is vital to propose a generic dashboard generation framework focusing on data visualization.

Mod2Dash is a software framework that enables users to capture their dashboard designs as models, which contain designs about the visual and functional aspects of dashboards. 
Mod2Dash automatically generates operational dashboards from these models by combining an extensible pool of dashboard widgets and deploying the resulting dashboard as a web application. 
As the auto-generated dashboards do not always strictly match the initial implicit design, Mod2Dash provides a GUI-driven customization approach that users can use to configure the dashboard according to their needs. 
The model also reflects the customization, resulting in a continuous feedback loop.
The proposed approach enables the users to rapidly prototype and deploy dashboards to validate and evaluate their designs.
Mod2Dash is also beneficial for researchers focusing on information visualization, as it enables them to create dashboard prototypes, generate dashboards, test on real stakeholder audiences, gather feedback, and improve dashboards in a fast and efficient manner.

We experimentally evaluated Mod2Dash's effectiveness based on a case study of cyber security visualization for decision support. 
We systematically collected, selected, and analyzed 31 cyber security dashboards from various sources of real-world scenarios.
A proof-of-concept of Mod2Dash was developed for modeling and reconstructing the selected 31 dashboards. Finally, we relied on a human-assisted study to count the matching design decisions between the original baseline dashboard and the dashboard constructed by Mod2Dash. The analysis and comparison of the counting on design decisions show a close matching between the pairs of dashboards, which means that Mod2Dash effectively captures dashboard designs and generates dashboards for real-world scenarios. Below are the main contributions of this paper:

\begin{itemize}

  \item We propose a novel and comprehensive dashboard language as a visualization model.
  \item We present an automated dashboard generation approach based on the proposed dashboard language, whereby different interpreters are designed to generate diverse dashboard designs.
  \item We present a GUI-driven customization approach, which allows a continuous feedback loop to improve the dashboard and reflect the dashboard design decisions to the model.
  \item We introduce the architecture design and a combination of open-source software technologies that can support such a dashboard framework and introduce the proof-of-concept of our proposed approach.

\end{itemize}

The rest of this paper is organized as follows. Section \ref{sec:relatedwork} outlines the review of related works regarding the visualization modeling language and automated dashboard generation. Section \ref{sec:mod2dash} presents the proposed Mod2Dash framework design. Section \ref{sec:implementation} introduces the detailed technologies that we used in the implementation. Section \ref{sec:evaluation}  describes the case study evaluation of Mod2Dash, and Section \ref{sec:conclusions} concludes our findings and future work.

\section{Related Work}\label{sec:relatedwork}

In this section, we discuss key researches related to visualization modeling languages and automated dashboard generation, which are essential fundamentals to our proposed approach for building an automated mechanism for creating dashboards.

\subsection{Visualization Modeling Language}

The data visualization languages can be regarded as either a manual or grammar that specifies how visualizations should be built and how they relate to data. 
Languages utilizing model-driven approaches facilitate the transition from hard-coding to models, thereby increasing productivity and scalability by describing critical aspects of a solution in a user-friendly manner and providing standardized templates for implementing the solution \cite{Sendall2003}.

Different programming languages have data visualization libraries or packages, for example, Matplotlib\footnote{Matplotlib: https://matplotlib.org/} for Python and ggplot2\footnote{ggplot2: https://ggplot2.tidyverse.org/} for R\footnote{R Programming Language: https://www.r-project.org/}. However, data visualization languages start from D3\footnote{D3: https://d3js.org/}. 
D3 \cite{Michael2011} allows users to capture their visualization designs and input data into arbitrary document elements. 
A dynamic transforming mechanism generates and modifies data visualization interfaces by combining visualization and interaction techniques.

While D3 is a supporting layer better suited for novel design ideas, Vega is a higher-level visualization specification language on top of D3 \cite{vegaandd3}. 
Vega \cite{Satyanarayan2016} provides a system architecture focused on declarative visual and interactive design for data visualization.
Vega implements a reusable and shareable chart component that automatically generates customizable and programmatic visualizations \cite{vegaandd3}.

Vega-Lite\footnote{Vega-Lite: https://vega.github.io/vega-lite/} \cite{Satyanarayan2017} is a subset and extension of Vega. It provides graphics and visual encoding rules, the compiler ingests a portable JSON syntax, and generates low-level Vega specifications. Especially, Vega-Lite introduces \textit{selections} concept in the specifications and parameterizes visual encoding into interactive features.
The publication of Vega-Lite is a milestone in the data visualization area. It facilities the development of research areas such as visualization recommendation, infographics authoring, and image reverse-engineering. 
For example, Data2Vis \cite{Dibia2018a} formulates visualization recommendation as a language translation problem. It recommends visualization by mapping and translating data specifications to visualization specifications in the Vega-Lite language. Poco and Heer \cite{Poco2017} investigate transforming a chart image into Vega-Lite-like visual encoding in reverse-engineering research.

The above languages \cite{Michael2011, Satyanarayan2016, Satyanarayan2017} focus on chart-level visualization or accumulating several different charts. 
They capture the design decisions in a single chart or interact with several associated charts. However, they are not capable of more design elements at a dashboard level, such as chart locations and dashboard-level interactions.
Moreover, the coming era of big data requires presenting and visualizing knowledge from a massive volume of data \cite{Ullah2022}. A single chart or few charts cannot present the whole picture of the knowledge. Therefore, a more comprehensive dashboard-level visualization language is important to present complex data and relationships.

\subsection{Automated Dashboard Generation}

\begin{table}[h]\caption{\label{tab:automation}Related work in model-driven dashboard generation.}
\begin{center}
\begin{tabular}{p{0.15\linewidth}|ccccc}
\hline
Reference                    & Discipline                                                    & Layout           & \begin{tabular}[c]{@{}c@{}}Chart\\ Interaction\end{tabular}  & \begin{tabular}[c]{@{}c@{}}Widget\\ Interaction\end{tabular}  & Customization \\ \hline
Palpanas et al.\cite{Chowdhary2006, Palpanas2007}             & Business    &   \begin{tabular}[c]{@{}c@{}}Hard coded\\ Template\end{tabular}    & \pmb{\ding{53}}& \pmb{\ding{53}}    & Model-based      \\ \hline
Kintz et al.\cite{Kintz2012, Kintz2017}               & Business &   \begin{tabular}[c]{@{}c@{}}Hard coded\\ Template\end{tabular}    & \pmb{$\checkmark$}        & \pmb{\ding{53}} & Model-based \\ \hline
Vazquez-Ingelmo et al.\cite{Vazquez-Ingelmo2019, Vazquez-Ingelmo2020, Vazquez-Ingelmo2020a}     &    \begin{tabular}[c]{@{}c@{}}Knowledge/\\ Interactions\end{tabular}                     &  Widget Size & \pmb{$\checkmark$}  & \pmb{\ding{53}}      & Model-based \\ \hline
Mod2Dash      &        Generic                           &  \begin{tabular}[c]{@{}c@{}}Widget Size\\ Location\end{tabular} & \pmb{$\checkmark$} & \pmb{$\checkmark$}  & \begin{tabular}[c]{@{}c@{}c@{}}Model-based\\GUI-based\\ More than 22 features\end{tabular}       \\ \hline
\end{tabular}
\end{center}
\end{table}

The automated dashboard generation process is the mechanism that generates dashboards automatically based on the rules. The research on this topic can be divided into data automation and visualization automation.
Data automation focuses on data aggregation and data preparation. For example, the solution implemented by Riege \cite{Riege2019} presents automated data filtering techniques in manufacturing sensor environments to find the most critical data and filter out the out-of-control data before visualizing it in a dashboard. Garabet et al. \cite{Topalian-Rivas2020} propose a module concept in the manufacturing environments to deal with packaging information before being fed into a visualization template to fulfill dashboard automation.

However, the present work discusses automation in visualization generation, especially the model-driven dashboard visualization frameworks. Three series of research dedicated to creating dashboards upon dashboard languages, as shown and compared in Table~\ref{tab:automation}. 
Chowdhary et al. \cite{Chowdhary2006} and Palpanas et al. \cite{Palpanas2007} use the XML meta-model representation to define a dashboard for modeling and visualizing business performance. Their model captures user roles, data metrics,  user interface templates, pages, and menus. They also map the model with hardcoded and predefined report templates. The code generator is designed to be modified in the deployment environment and generates deployable software components, and the final dashboard application is deployed on a J2EE application server. 
The approach proposed by Kintz et al. \cite{Kintz2012, Kintz2017} also uses XML to describe a dashboard for monitoring and controlling business processes. Kintz \cite{Kintz2012} captures generic dashboard information, dashboard visualization elements, alerting, interactions, and data. Kintz et al. \cite{Kintz2017} expand the model by adding roles and views into the meta-model. They implement the visualization mechanism in a web browser-based JavaScript interface.

Vazquez-Ingelmo et al. \cite{Vazquez-Ingelmo2019, Vazquez-Ingelmo2020, Vazquez-Ingelmo2020a} conduct a series of studies and also proved the meta-model's usability in different disciplines when building a dashboard. For example, in \cite{Vazquez-Ingelmo2019}, they propose a dashboard meta-model that mainly elaborates the user, the size of the container, and the components in a healthcare knowledge management case. In \cite{Vazquez-Ingelmo2020}, the authors model interaction patterns by using XML specifications in a learning analytics case. A proof-of-concept in \cite{Vazquez-Ingelmo2020a} is developed utilizing the Software Product Line generation process and based on Vega-Lite chart language.

As shown in Table~\ref{tab:automation}, among these dashboard languages, Palpanas et al. \cite{Chowdhary2006, Palpanas2007} and Kintz et al. \cite{Kintz2012, Kintz2017} propose the dashboard meta-model for monitoring business performance, while the approaches proposed by Vazquez-Ingelmo et al. \cite{Vazquez-Ingelmo2019, Vazquez-Ingelmo2020, Vazquez-Ingelmo2020a} focus on integrating the dashboard meta-model with other meta-models (e.g., knowledge management in learning ecosystem and interactions). However, these dashboard languages have limited generalizability to other disciplines due to the design constraints for the particular use cases. Therefore, the Mod2Dash dashboard language will target a more generic dashboard visualization language and be designed as a pure visualization language for real-world cross-discipline dashboard generation.

All these dashboard languages fail to represent the dashboard layout and widget location. Palpanas et al. \cite{Palpanas2007} and Kintz et al. \cite{Kintz2017} map the model with pre-defined templates to generate the dashboard. Vazquez-Ingelmo et al. \cite{Vazquez-Ingelmo2019, Vazquez-Ingelmo2020, Vazquez-Ingelmo2020a} design the widget size but fail to consider organizing the widget location in their meta-model. 

Two of these approaches rule the chart-level interactions in the model. For example, Kintz et al. \cite{Kintz2017} provide interaction capabilities such as \textit{Zoom}, \textit{Pan}, and \textit{Drill down}. Vazquez-Ingelmo et al. \cite{Vazquez-Ingelmo2019} also mention the \textit{Drill down} interaction design in their model. However, their designs stay at chart-level interaction, and they fail to represent more comprehensive interactions at the dashboard level, which we will discuss in Section \ref{subsec:language}.

The models rule how the personalized visualization will be presented, so the diversity of customization functions mainly depends on the model design.
All the dashboard languages mentioned above provide a limited number of model-based customization and do not attempt to provide reversed GUI-based customization. However, Mod2Dash provides more than 22 customization features and provides a GUI-based customization approach, which will be discussed in Section \ref{subsec:customization}.

\section{Mod2Dash} \label{sec:mod2dash}

In this section, we first present the preliminary of our approach, including the motivating example, the approach workflow, and the usage scenario. Then we introduce three main components of our approach in detail.

\subsection{Preliminary}

\subsubsection{Motivating Example}

\begin{figure}[h]
\begin{minipage}[h]{0.98\textwidth}
\centering
\includegraphics[width=1\textwidth]{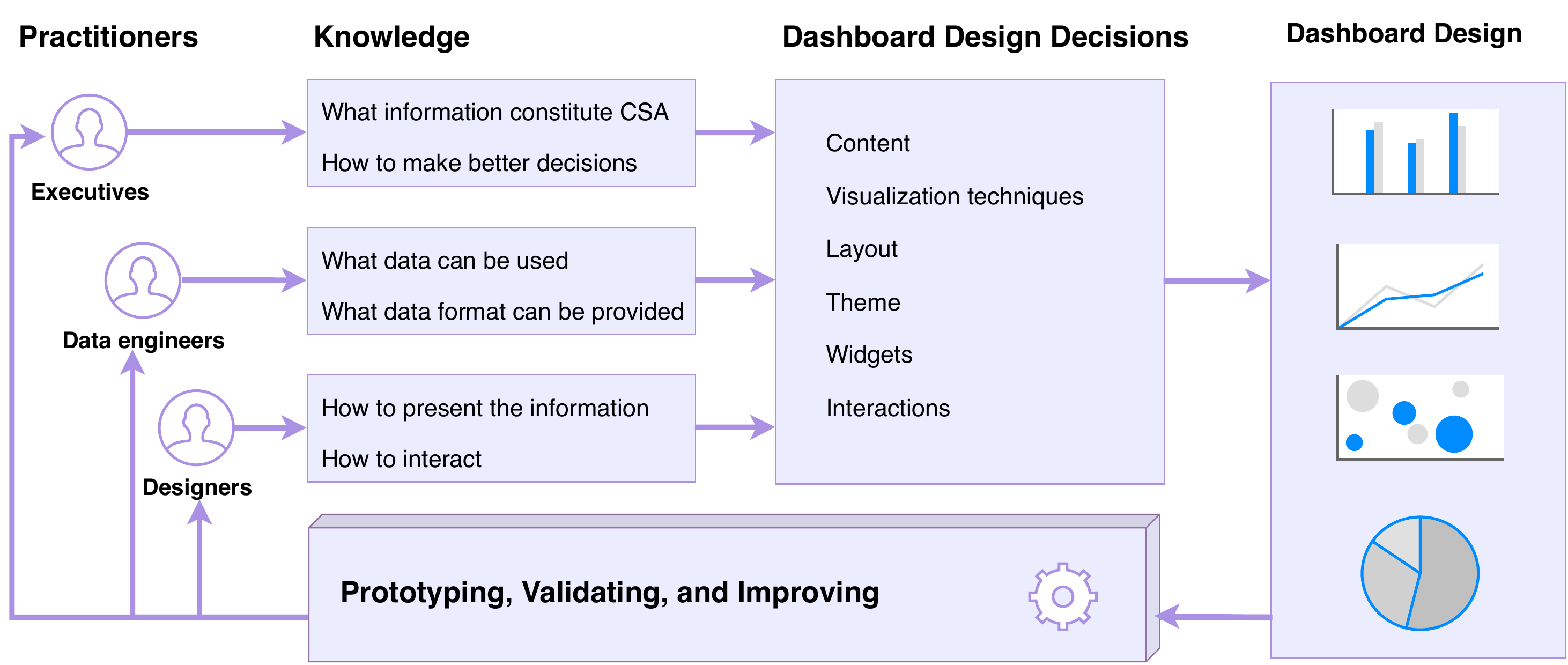}
\caption{Cyber security dashboard prototyping and validating.}
\label{motivating}
\end{minipage}\hfill%
\end{figure}

Figure \ref{motivating} shows a motivating example of our approach, which is a dashboard prototyping and validating project in the cyber security domain.

In this example, the dashboard practitioners are executives, data engineers, and user interface (UI) designers. The executives know what information is helpful for their CSA and decision-making, and they are the final consumers of the dashboard. The data engineers have organizational CSA knowledge and what data can be offered and fed into the dashboard. Finally, the UI designers analyze the dashboard requirements, design the visual and interactive effects, and deliver to the development and validation phases. All of them contribute to dashboard design decisions; however,  practitioners sometimes change designs rapidly or significantly to produce a final dashboard. During the development process, they need to communicate and version-control the changes.

As the example illustrates, we may be able to create a unified language to express dashboard design decisions made by all parties and develop an automated dashboard composition process to translate that language into an operational dashboard. Consequently, the dashboard practitioners can version control their designs, quickly prototype their designs, and collaborate on improving their designs.

\subsubsection{Workflow and Usage Scenario}
Mod2Dash is designed to address the challenges faced by dashboard practitioners when capturing their design decisions, transforming their design decisions into real dashboards, and improving the dashboard designs.

\begin{figure}[h]
\begin{minipage}[h]{0.98\textwidth}
\centering
\includegraphics[width=1\textwidth]{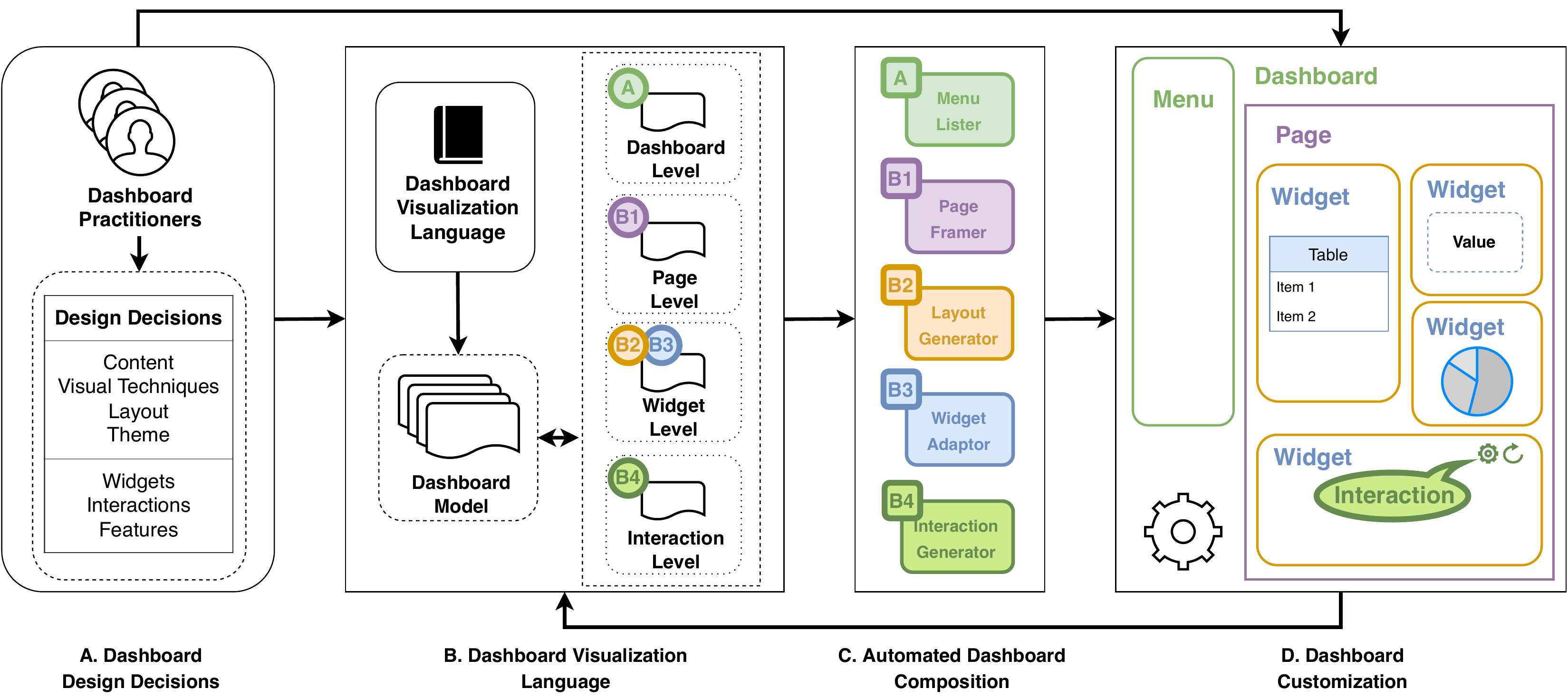}
\caption{Mod2Dash overall workflow. Dashboard practitioners' implicit design decisions are captured in the dashboard visualization model by following the dashboard visualization language grammar. Next, the model is fed into the automated dashboard composition mechanism to generate an operational dashboard. The dashboard designs can then be fine-tuned through the GUI-based customization approach, and the updated dashboard model will also reflect the changes and be stored.}
\label{workflow}
\end{minipage}\hfill%
\end{figure}

Mod2Dash proposes a meta-model, which describes the possible entities and relationships used in the concrete dashboard models, and it allows dashboard practitioners to encapsulate their implicit design decisions in a dashboard model. A dashboard generation mechanism is also designed to automate the process of turning the dashboard model into an actual dashboard to assist dashboard practitioners in prototyping their designs and communicating with others. Moreover, a GUI-based customization capacity is also proposed for practitioners continuously improve dashboard designs. 
The overall workflow of Mod2Dash is shown in Figure ~\ref{workflow}. Its usage scenarios are as follows: 

First, dashboard practitioners capture the dashboard design decisions (e.g., visualization techniques, dashboard layout, theme, widgets, and interactions) into a dashboard model, and this process can be done manually, semi-automatically, or automatically by following the proposed dashboard language. 
If the dashboard practitioner has coding experience, she can write the model from the beginning or use other programming languages (e.g., Python, Java, and JavaScript) to generate or improve a dashboard model in a large-scale and automatic setting by following the proposed dashboard language (see Section \ref{subsec:language}).
If the practitioner is unfamiliar with coding, she may construct this dashboard model by dragging and dropping the widget in the GUI customization mode. Then the dashboard model can be generated accordingly and stored (see Section \ref{subsec:customization}).

Second, the visualization model is directly fed to the automated dashboard composition mechanism, and the mechanism can build a web-based operational dashboard from the model. The user can import their pre-defined dashboard model into the mechanism for quickly prototyping her dashboard design, and her team members can see the design, give feedback, or further improve the design based on the existing model.

Then, the GUI-driven customization approach allows the practitioner to fine-tune the dashboard designs. For example, if the dashboard is built based on the hand-coded or script-generated visualization model, it may not fulfill the design requirements; the practitioner can manually adjust the elements in the dashboard via the GUI customization mode. 

Finally, the improved dashboard design decisions are stored as the updated visualization model. So the feedback loop allows the practitioners to improve the dashboard in iteration.

The following subsections discuss how the Mod2Dash approach is designed in detail.

\subsection{Dashboard Visualization Language} \label{subsec:language}

The dashboard visualization language serves as a grammar that describes how the visualization should be presented and the skeleton upon which dashboard generation can be automated.
The dashboard language can help dashboard practitioners capture their designs into a model so that the designs can be version-controlled, backed up, and collaborated among teams.
From the perspective of software engineering, the model-driven approach facilitates the transformation of the software development process from hard-coding to modeling. The productivity and scalability of the development can be increased by using commonly applicable templates and describing essential aspects of the solution in a human-friendly manner \cite{Sendall2003}. 

This subsection presents our process for designing the dashboard model and the detail of the proposed language.

\subsubsection{Dashboard Prototyping Design}

\begin{figure}[h]
\begin{minipage}[b]{0.29\textwidth}
\centering
\includegraphics[width=1\textwidth]{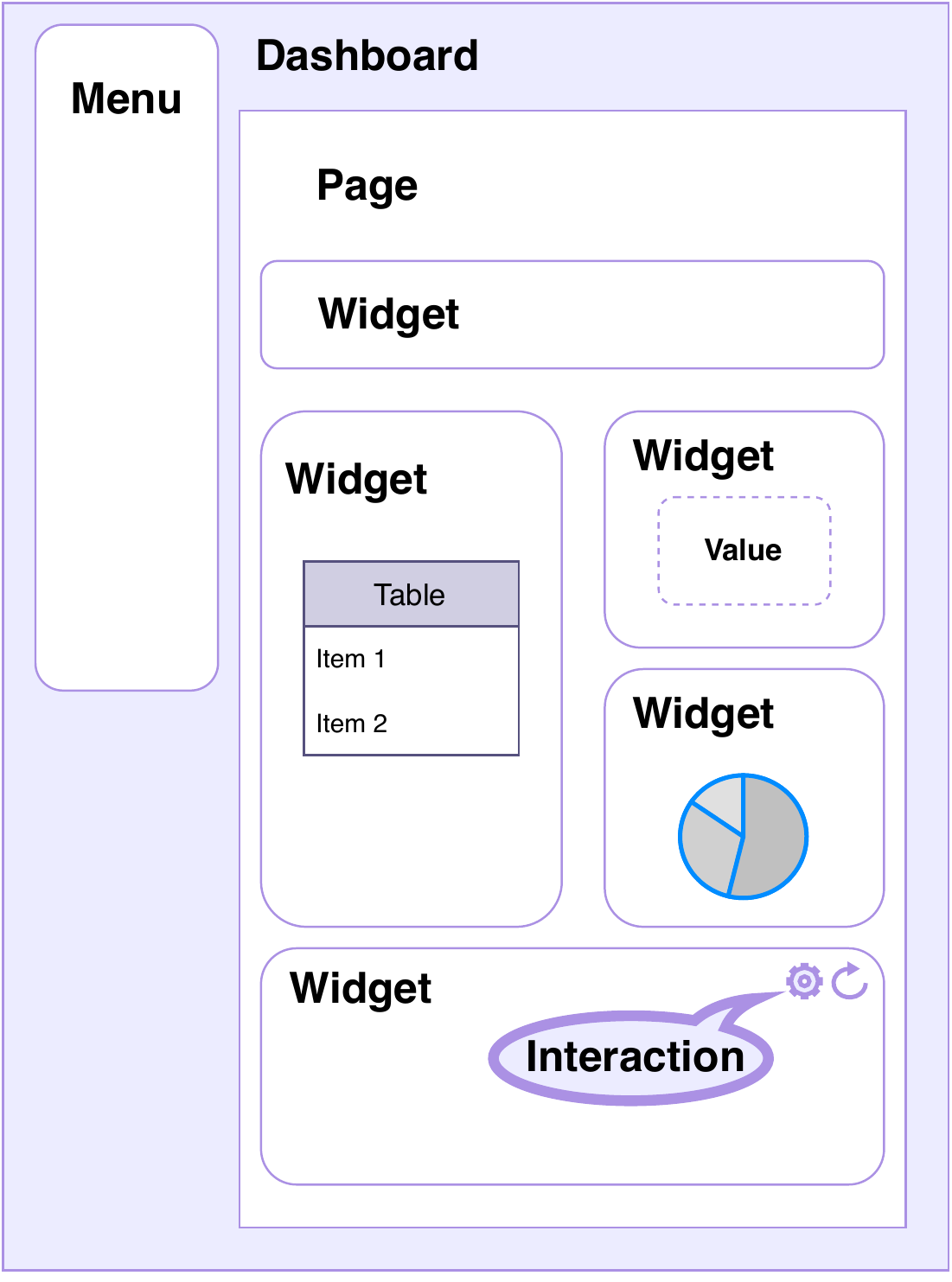}
\caption{The dashboard frame prototyping design.}
\label{dashboardlayout}
\end{minipage}\hfill%
\begin{minipage}[b]{0.7\textwidth}
\centering
\includegraphics[width=1\textwidth]{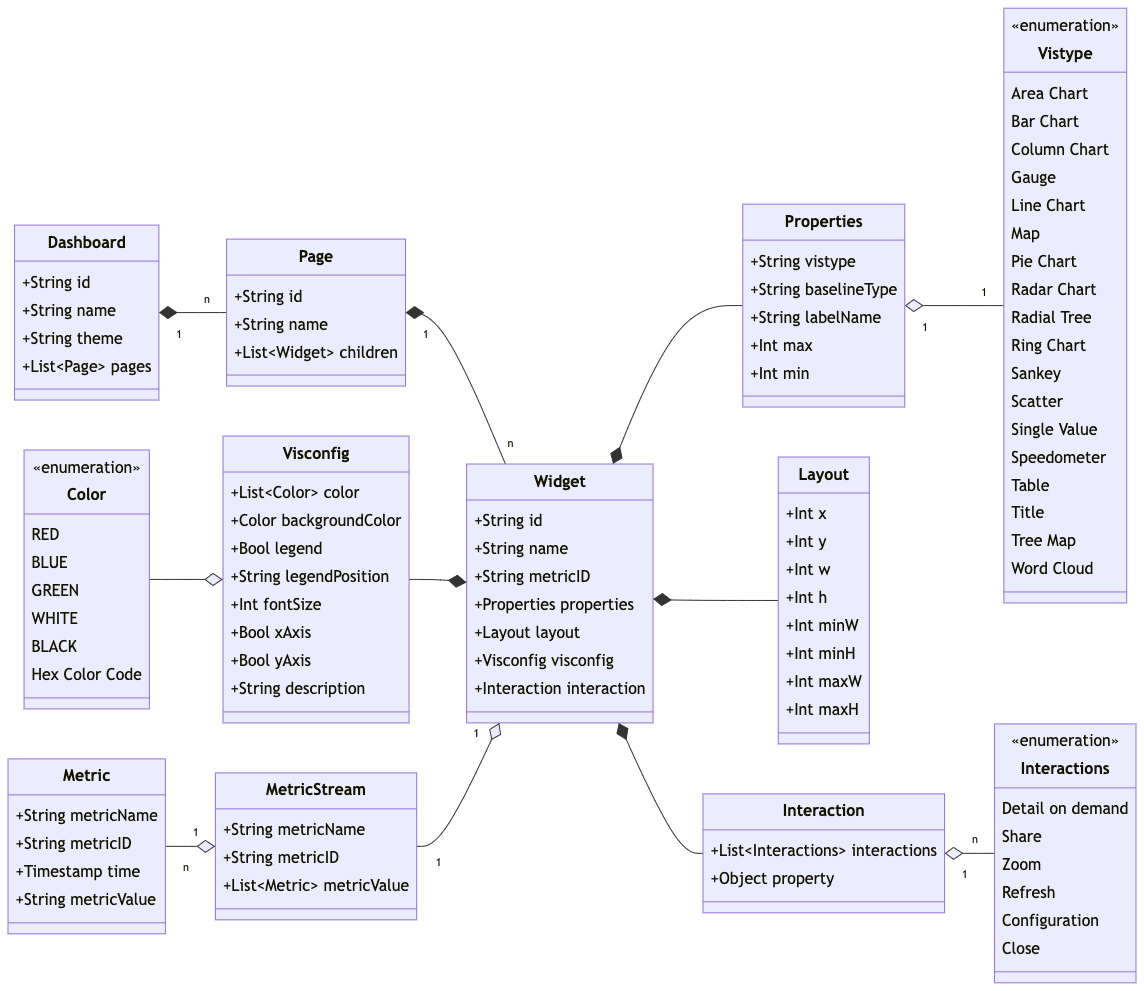}
\caption{The meta-model of Mod2Dash dashboard visualization language.}
\label{language}
\end{minipage}\hfill%

\end{figure}

The dashboard is not simply the accumulation of multiple charts, and it involves many design aspects. 
Therefore, to make the dashboard visualization language reflect and be capable with the real-world dashboards, we firstly examined the dashboard designs from the \textit{"Information Dashboard Design"} book by Few \cite{Stephen} and the dashboards collected by Sarikaya et al. \cite{Sarikaya2019}, we then conceptualize a dashboard prototyping design. As we can see in Figure ~\ref{dashboardlayout}, a dashboard should have the main content page and a menu for navigating between the different pages. Each widget on a page could be different in size and is composed of different visual representation techniques (e.g., charts, tables, and values). 

\subsubsection{Meta-Model}

We get inspired further by previous model-driven automated dashboard generation research \cite{Chowdhary2006, Kintz2012, Vazquez-Ingelmo2019, Vazquez-Ingelmo2020} and chart-level data visualization language research \cite{Michael2011, Satyanarayan2016, Satyanarayan2017}. According to the prototyping design shown in Figure \ref{dashboardlayout}, the dashboard visualization language should consist of three main elements: the components, the layout, and the functions. The components are the visual elements such as menu, widgets, and charts, including their visual properties such as color, font size, and axis. The layout is how these components are located and organized in a dashboard. And the functions include the interactions and the data connections.

Considering all these ideas, we design a four-level hierarchy meta-model structure for capturing the generic properties and configurations of dashboard representations. Figure \ref{language} shows the meta-model of the proposed dashboard visualization language. Here we discuss each hierarchy in detail:

\noindent\begin{minipage}[]{.5\textwidth}

\begin{lstlisting}[language=json,firstnumber=1,label={lst:basicmodel},caption={A simple dashboard model example in JSON.}]
{
    "id": "Dashboard_Sample",
    "name": "Sample Dashboard",
    "theme": "light",
    "pages": [
        {
            "id": "0",
            "name": "Sample Page",
            "widgets": [
                {
                    "id": "p0-i0",
                    "properties": {
                        "vistype": "title",
                        "title": "Sample Title Widget"
                    },
                    "layout": {
                        "w": 4,
                        "h": 2,
                        "x": 0,
                        "y": 0
                    }
                },
                {
                    "id": "p0-i1",
                    "name": "Sample Pie Widget",
                    "properties": {
                        "vistype": "pie",
                        "childrenname": [
                            "Value1",
                            "Value2",
                            "Value3"
                        ]
                    },
                    "layout": {
                        "w": 4,
                        "h": 8,
                        "x": 0,
                        "y": 2
                    },
                    "visconfig": {
                        "colour": [
                            "#82b365",
                            "#9673a6",
                            "#6c8ec0"
                        ]
                    },
                    "interaction": {
                        "interactions": [
                            "Detail on demand"
                        ],
                        "detail": {
                            "target": "0",
                            "method": "pure"
                        }
                    }
                }
            ]
        }
    ]
}
\end{lstlisting}
\end{minipage}\hfill
\begin{minipage}[]{.49\textwidth}

\begin{figure}[H]
              \includegraphics[width=\linewidth]{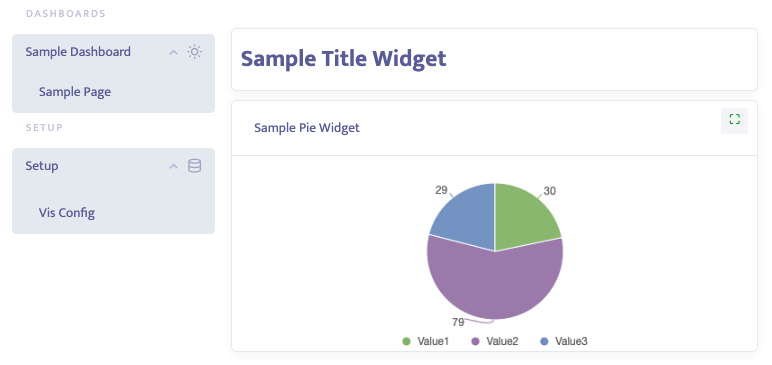}
              \caption{The dashboard generated from the simple dashboard model in Listing \ref{lst:basicmodel}. In this example, the dashboard is in a light theme and includes one page. The page includes two widgets, one is a title widget, and another one is a pie chart widget. The pie chart has three values with three colors, and the interaction in this widget is Detail on demand.}
              \label{basicdashboard}
          \end{figure}
\end{minipage}

\begin{itemize}
  \item \includegraphics[height=2ex]{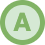} \textbf{Dashboard} is the frame of the web application. \textit{Dashboard} is the top hierarchy of the model, the class \textit{Dashboard} includes parameters: dashboard ID, dashboard name, theme, and children pages. The name of \textit{Dashboard} is the title of the web application, \textit{Dashboard} consists of more than one \textit{Pages}.

  \item \includegraphics[height=2ex]{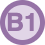} \textbf{Page} is the main content in a dashboards, consisting a group of \textit{Widgets}. The \textit{Page} class includes parameters: page ID, page name, and children widgets. The names of the \textit{Page} are used to generate the navigation menu, and different pages can be navigated from the menu or \textit{detail on demand} interaction feature.
  
  \item \includegraphics[height=2ex]{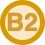}\includegraphics[height=2ex]{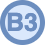} \textbf{Widget} is the basic element in the visualization. It frames the chart, text, or table. It is named container, chart, or panel in other research or products. In the dashboard model, the widget level hierarchy describes where the widget is located and how the widget is visualized. It contains parameters: widget ID, widget name, metric ID, interactions, and a series of optional objects such as visualization properties, layout, and visualization configuration.
  
  \item \includegraphics[height=2ex]{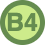} \textbf{Interactions} at the dashboard level and widget level are different from the interactions provided by the chart. Dashboards provide more interactive features than the built-in interactions of a chart library. 
  Therefore, this paper focuses on the dashboard and widget interactions rather than the interactions provided at the chart level. The \textit{Interaction} class describes what interactions will be used for a particular widget and detailed configuration properties for the interaction.
  
\end{itemize}

\begin{figure}[b]
\includegraphics[width=0.98\textwidth]{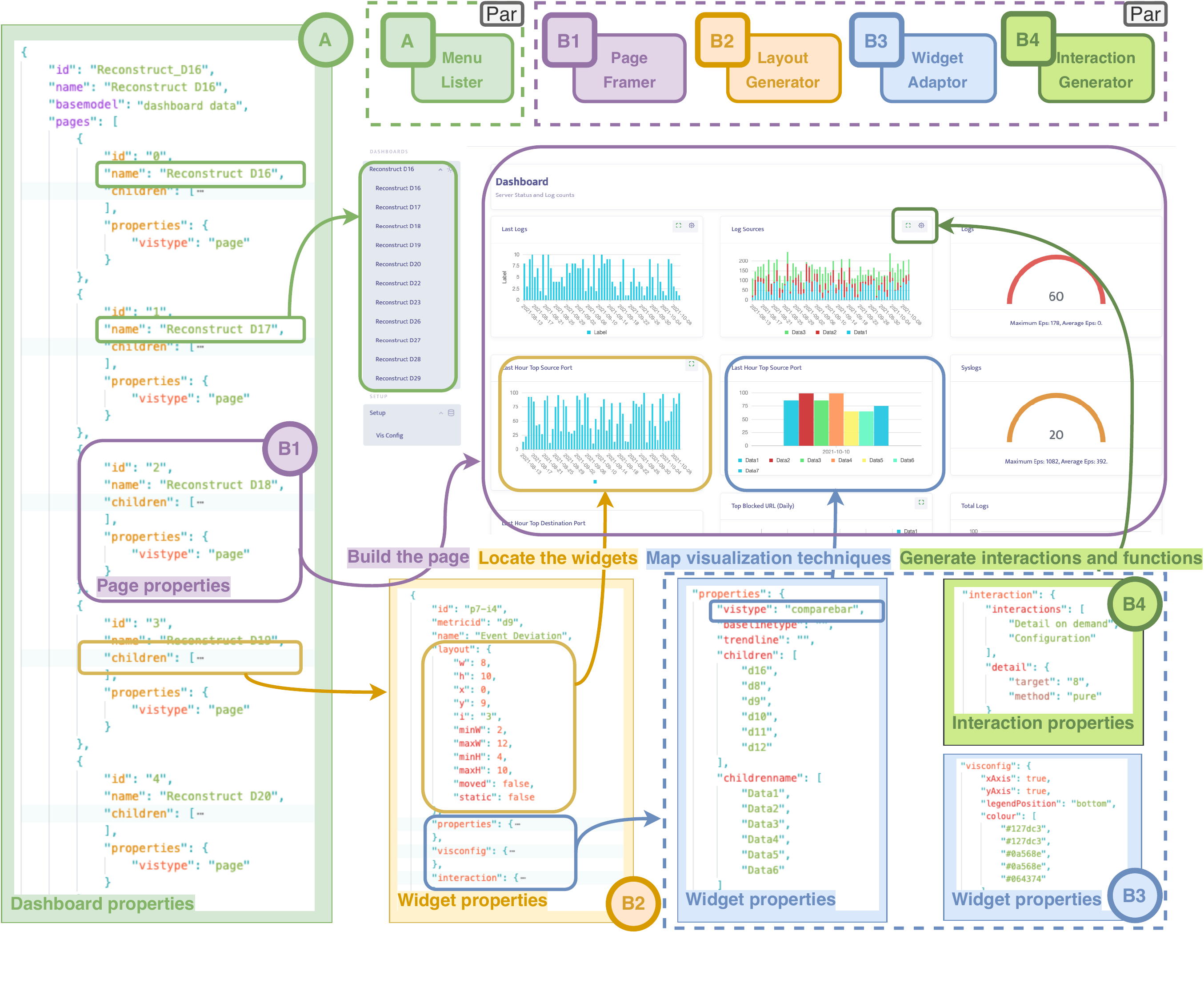}
\caption{Model example of a cyber security dashboard, including the four-level hierarchy models. \includegraphics[height=2.3ex]{image/icon/A1.png} is a sample model structure of dashboard-level properties, \includegraphics[height=2.3ex]{image/icon/B11.png} is the configuration of a page in the dashboard, \includegraphics[height=2.3ex]{image/icon/B21.png} and \includegraphics[height=2.3ex]{image/icon/B31.png} are the sample model of widget-level properties (i.e., locations and visual configurations), and \includegraphics[height=2.3ex]{image/icon/B41.png} is the specification of the interactions.}
\label{composition}
\end{figure}

\subsubsection{Dashboard Model}

We adopt JSON\footnote{JSON: https://www.json.org/} (a JavaScript Object Notation) to represent a dashboard specification. JSON is a commonly used data-interchange language, and it is supported in most major programming languages \cite{wiki:json}. We can formulate a dashboard representation by following the JSON language syntax and meta-model design mentioned above. Listing \ref{lst:basicmodel} shows a simple dashboard model example, the dashboard contains two widgets on one page, and the corresponding dashboard is shown in Figure \ref{basicdashboard}.

Thus, the dashboard practitioners can capture their dashboard designs into a model according to their real-world dashboard design requirements. 
Figure \ref{composition} shows another more complex model example from a cyber security dashboard. This example highlights the four-level hierarchy meta-model structure design.

\subsection{Automated Dashboard Composition}

When the dashboard visualization model has been created, the automated dashboard generation mechanism allows the dashboard practitioners' design decisions to be presented instantly to effectively communicate and validate the dashboard design.

\begin{figure}[h]
\includegraphics[width=0.98\textwidth]{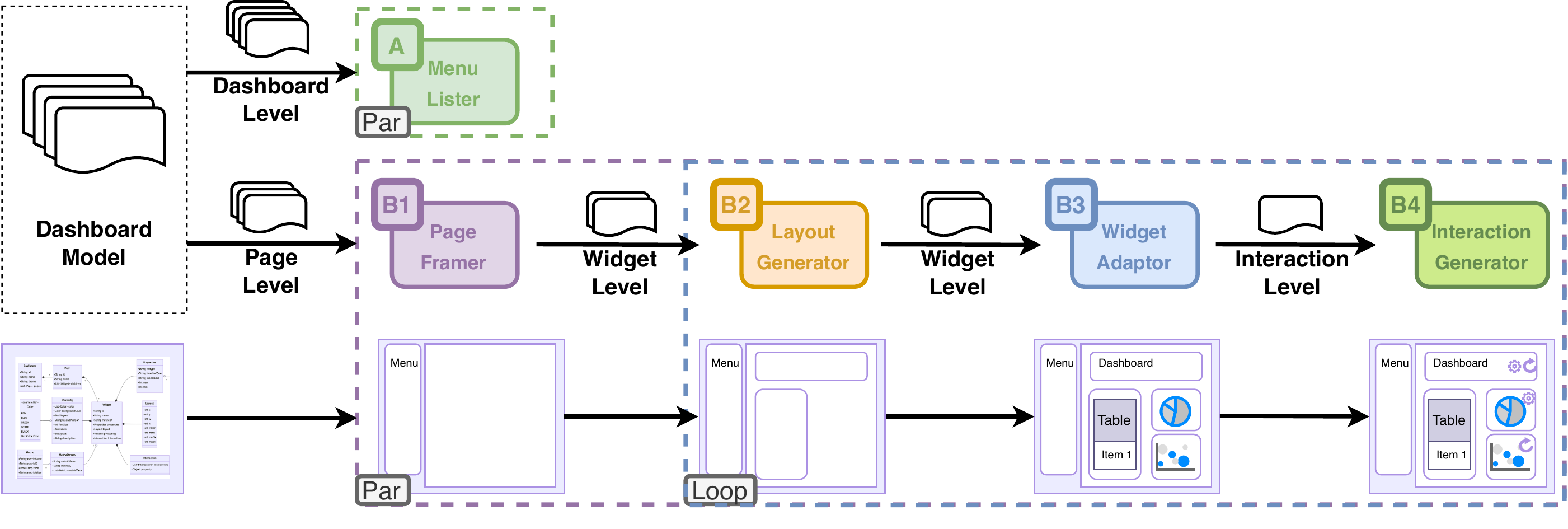}
\caption{Mod2Dash automated dashboard composition workflow. \includegraphics[height=2ex]{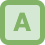} takes the whole dashboard-level model, extracts pages names, and generates the menu for the dashboard. \includegraphics[height=2ex]{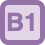} uses page-level model to frame a page. \includegraphics[height=2ex]{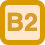} uses the widget-level model to locate each widget one by one, and \includegraphics[height=2ex]{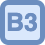} also uses the widget-level model to render different visualization techniques. Finally, \includegraphics[height=2ex]{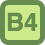} takes the interaction-level model to generate the interaction features. In the sequence of renderings, \includegraphics[height=2ex]{image/icon/A2.png} and the series of  \includegraphics[height=2ex]{image/icon/B12.png} \includegraphics[height=2ex]{image/icon/B22.png} \includegraphics[height=2ex]{image/icon/B32.png} \includegraphics[height=2ex]{image/icon/B42.png} are rendered in parallel.}
\label{compositionflow}
\end{figure}

Figure \ref{compositionflow} shows the overview of the dashboard composition mechanism, especially how the dashboard model flows in different composition interpreters.
Below we describe the dashboard composition interpreters and the automated rendering procedures in detail.

\begin{itemize}
    \item \includegraphics[height=2ex]{image/icon/A2.png} \textbf{Menu Lister} reads and extracts the page names from the model and generates a clickable menu at the top left of the dashboard. When the user clicks an item in the menu, the \textit{Page Framer} will be triggered to render the corresponding dashboard page.
    \item \includegraphics[height=2ex]{image/icon/B12.png} \textbf{Page Framer} reads the current page number from the URL when a page is navigated, and it also reads the page-level model to generate the page. The \textit{Layout Generator} will be instantiated in \textit{Page Frame} with the widget-level parameters. The process happens synchronously with the \textit{Menu Lister}.
    \item \includegraphics[height=2ex]{image/icon/B22.png} \textbf{Layout Generator} reads widget layout properties in widget-level models, which are passed from the \textit{Page Framer}, and allocates each widget at the corresponding location. Each location will instantiate a \textit{Widget Adaptor} to render a chart component. This step builds up the skeleton of the dashboard.
    \item \includegraphics[height=2ex]{image/icon/B32.png} \textbf{Widget Adaptor} is a basic interpreter to render widget-level design decisions. It maps different visualization types into web-based widget components according to the widget-level model. The individual widget component interprets detailed configurations, such as the composition of metrics, color combinations, legend settings, and axis settings. If the interactions exist in the current widget, the \textit{Interaction Generator} will be instantiated by passing the interaction-level model.
    \item \includegraphics[height=2ex]{image/icon/B42.png} \textbf{Interaction Generator} instantiates interactions and triggering functions based on the detailed interaction configuration parameters. For example, if the interaction type is \textit{Detail on demand}, \textit{Interaction Generator} will render an icon representing the feature; if the user clicks it, the corresponding function will be triggered and jump to a new page.

\end{itemize}

Figure~\ref{composition} shows the detailed dashboard composition mechanism for the aforementioned cyber security dashboard model; it details the procedures of the proposed approach translating the sample model into a web-based dashboard. 
\includegraphics[height=2ex]{image/icon/A2.png} \textit{Menu Lister} uses the dashboard-level model \includegraphics[height=2ex]{image/icon/A1.png} and generates the navigation menu. Next, \includegraphics[height=2ex]{image/icon/B12.png} \textit{Page Framer} takes the page-level model \includegraphics[height=2ex]{image/icon/B11.png} to render the current page. Then \includegraphics[height=2ex]{image/icon/B22.png} \textit{Layout Generator} reads the widget-level model, especially location properties \includegraphics[height=2ex]{image/icon/B21.png}, to locate each widget, and \includegraphics[height=2ex]{image/icon/B32.png} \textit{Widget Adaptor} reads the widget-level visual properties \includegraphics[height=2ex]{image/icon/B31.png} to map different visualization configurations with widget components. Finally, \includegraphics[height=2ex]{image/icon/B42.png} \textit{Interaction Generator} reads the interaction-level model \includegraphics[height=2ex]{image/icon/B41.png} to generate interactions and functions.

After the composition procedure above, a fully operational dashboard is built, and the resulting dashboard can be quickly presented to others, validate the design, and gather feedback to improve the dashboard design further.

\subsection{Dashboard Customization} \label{subsec:customization}

\textit{"The dashboard must be specifically to the requirements of a given person, group, or function"} \cite{Stephen}. 
The customization feature is essential for building a specific dashboard that helps users in a particular use case. The visualization configuration is also widely applied in dashboard practices \cite{logzio}.

\begin{figure}[h]
\begin{minipage}[h]{0.98\textwidth}
\centering
\includegraphics[width=1\textwidth]{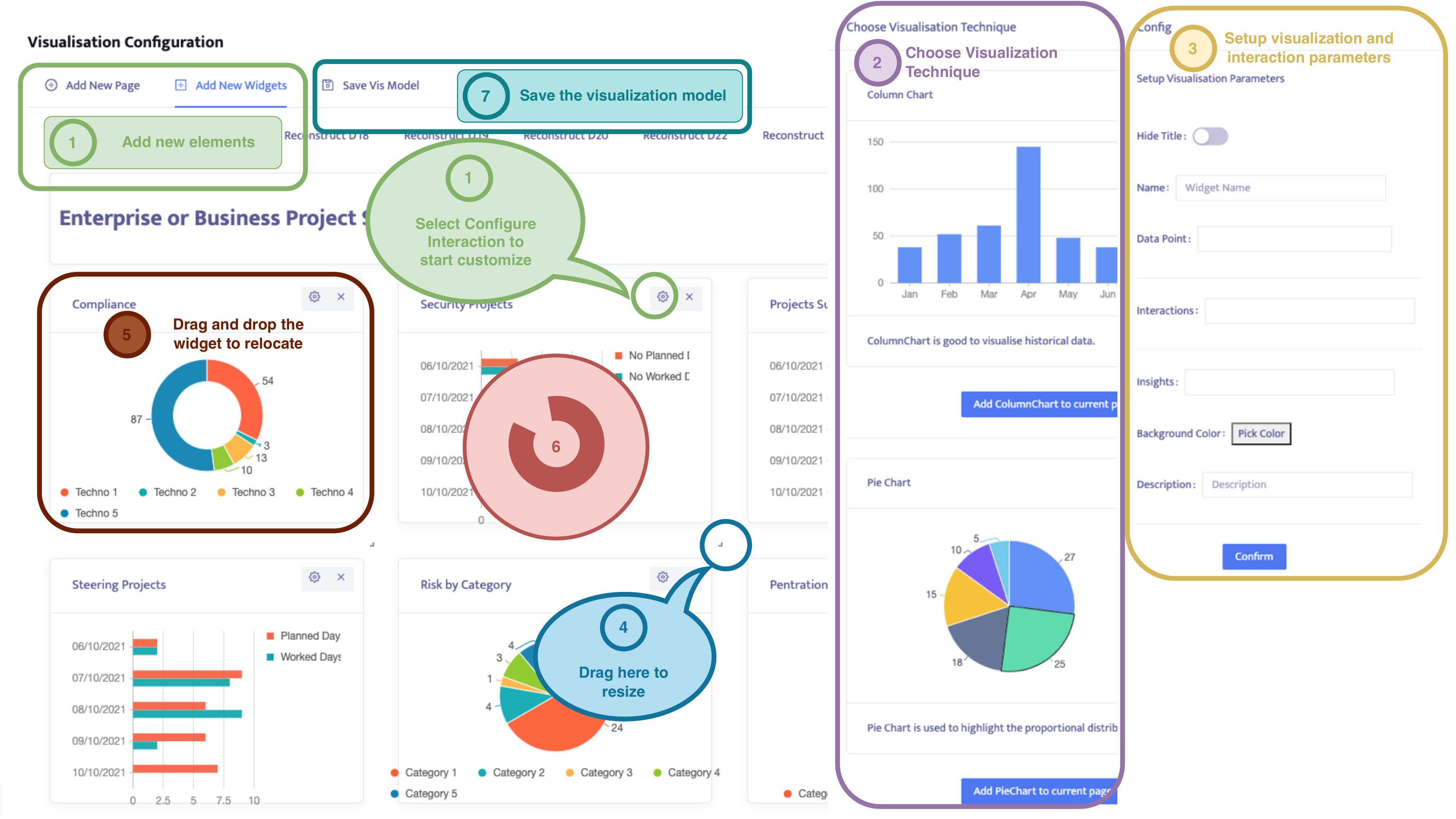}
\caption{Dashboard visualization building and configuring steps in the GUI.}
\label{GUI}
\end{minipage}\hfill%
\end{figure}

The actual dashboard produced by the auto-generation process from the initial model will never match the original design and the use case. Therefore, the GUI-driven customization approach allows dashboard users to tailor the dashboard to match their particular use cases, and the customization will also be reflected in the model, creating an improvement feedback loop.

\begin{table}[b]
\caption{\label{tab:configability}The customization features that Mod2Dash provides.}
\begin{tabular}{p{1.5cm}|p{2.5cm}|p{8.5cm}}
\hline
\multicolumn{2}{l|}{Configuration Item}           & Description                                                                                                         \\ \hline
\multirow{4}{*}{Dashboard}   & name                & The dashboard name is customizable, it shows as the title of the dashboard.                                         \\ \cline{2-3} 
                             & theme       & It includes dark and light themes, the user is able to choose from them.          
      \\ \cline{2-3} 
                             & base data model       & Base model is pointing to the data model.          
      \\ \cline{2-3} 
                             & switch model          & The user is able to switch between dashboard models.     
                             \\ \hline
\multirow{3}{*}{Page}        & new page            & Create a new page via GUI.                                                                                           \\ \cline{2-3} 
                             & name                & The page name is customizable, it is shown at the navigation menu.                                                  \\ \cline{2-3} 
                             & layout              & The layout of the page is customizable by drag and drop the widget.                                                 \\ \hline
\multirow{7}{*}{Widget}      & new widget                & Add a new widget to the page via GUI.                                                                       \\ \cline{2-3} 
                             & name                & The widget name is shown as the widget title.                                                                       \\ \cline{2-3} 
                             & visualization       & Many different visualization techniques can be chosen.                                                    \\ \cline{2-3} 
                             & resize              & The widget's size can be controlled by dragging the widget edge in the editing mode.                               \\ \cline{2-3} 
                             & move                & The widget can be relocate to a new location by dragging on the widget.                                             \\ \cline{2-3} 
                             & color               & The background color of the widget is customizable by a color picker.                                               \\ \cline{2-3} 
                             & metric id            & Which metric is showing in this widget can be selected, can be set with multiple metrics for some visualization.    \\ \hline
\multirow{6}{*}{\begin{tabular}[c]{@{}c@{}}Chart\\ Element\end{tabular}}     & layout              & The layout of the elements inside a widget is customizable.                                                         \\ \cline{2-3} 
                             & disable legend      & The legend can be hided.                                                                                            \\ \cline{2-3} 
                             & legend position     & The position of the legend in the chart is configurable.                                                            \\ \cline{2-3} 
                             & baseline            & The baseline for line chart and bar chart can be set to moving average or deviation.                                \\ \cline{2-3} 
                             & font size           & The font size of single value widget is configurable.                                                               \\ \cline{2-3} 
                             & disable axis label  & The label for x axis and y axis can be disabled.                                                                    \\ \hline
\multirow{2}{*}{Interaction} & select interactions & The interaction is configurable for a widget, such as detail on demand, share, zoom, refresh, configuration, close. \\ \cline{2-3} 
                             & configure interactions & The properties that when triggering an interaction are configurable.                                                    \\ \hline
\end{tabular}
\end{table}

Mod2Dash enables the user to choose from different visualization techniques, drag and drop to move widgets around, drag to resize the widgets, and manipulate the element color in a chart. 
Besides these basic visualization configurations,
Mod2Dash provides configuration features in different components or levels, even the ability to configure the layout of the elements inside a widget, which is benefited from the four-level hierarchy dashboard language design.
Moreover, Mod2Dash provides configurations on interactions. 
Table ~\ref{tab:configability} shows the customization features that Mod2Dash offers.

The customization features rely on a GUI implementation, and users can create new elements or configure the existing elements of the dashboard through the GUI in the configuration mode. We describe the building and configuring procedure below, and Figure ~\ref{GUI} also shows this procedure.

\begin{itemize}
    \item \textbf{\textit{Step1:}} Start by adding new pages or widgets on the toolkit or selecting the configuration interaction icon on the widget. The GUI will pop out a side window for configuration.
    \item \textbf{\textit{Step2:}} If choosing to add a new widget or edit a widget, the first level side window allows users to choose visualization techniques. After selecting a visualization technique, the GUI will pop out the second-level side window.
    \item \textbf{\textit{Step3:}} If choosing to add a new page or select the visualization technique, the second level side window allows users to set up visualization properties and interaction parameters. After choosing and typing in the required information, a corresponding widget will be rendered on the page, and the new widget will be automatically inserted at the bottom of the current page.
    \item \textbf{\textit{Step4:}} In the configuration mode, the widget is resizable, and the user can drag the resize icon on a widget to resize the widget. The content inside a widget frame is also adaptable based on the widget size.
    \item \textbf{\textit{Step5:}} In the configuration mode, the widget is also draggable to relocate.
    \item \textbf{\textit{Step6:}} Redo the \textit{Step1} to \textit{Step5} to add or configure more widgets until fulfill the dashboard design requirements.
    \item \textbf{\textit{Step7:}} The user's operation in each step results in a new model. Click the save button to save the visualization model to the database.
\end{itemize}

\section{Implementation} \label{sec:implementation}

Mod2Dash can be developed based on a combination of open-source software technologies. The deployment of Mod2Dash resembles a client-server database architecture, and all the components of the system reside on one computing host. 
This section describes the details of how Mod2Dash can be developed and deployed.

\subsection{Architecture} \label{subsec:architecture}

\begin{figure}[h]
\begin{minipage}[h]{0.88\textwidth}
\centering
\includegraphics[width=1\textwidth]{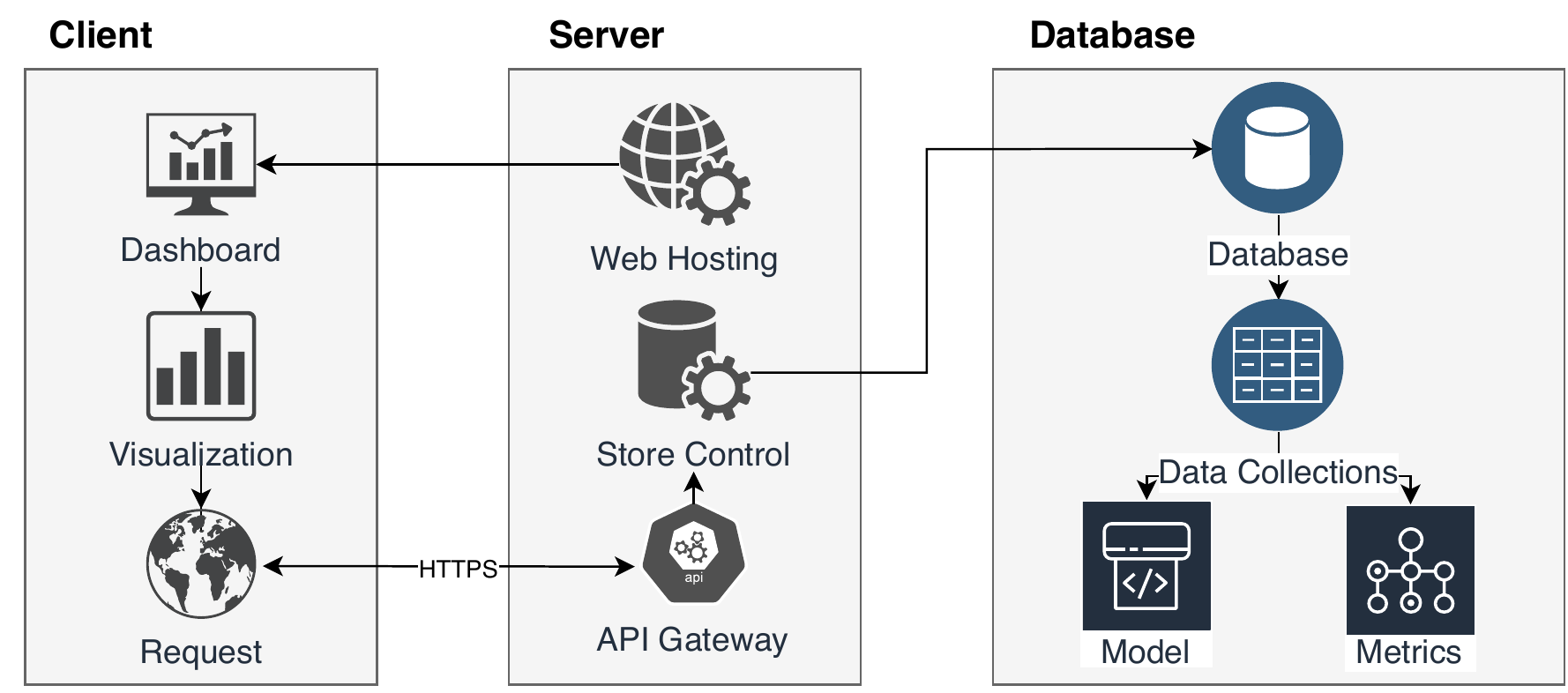}
\caption{Deployment architecture design of Mod2Dash.}
\label{architecture}
\end{minipage}\hfill%
\end{figure}

To support such a proposed mechanism, when considering the three core features of Mod2Dash design, we plan to implement it in a client-server database architecture \cite{Delis}. Figure ~\ref{architecture} shows the deployment architecture of Mod2Dash. 

The frontend web application is designed as a client-side application, and it involves the visual and interactive elements for the users. In this case, the frontend can generate dashboards based on different level interpreters, present the dashboard generation results, and customize the dashboard design.
The frontend runs as a web service by the \textit{Web Hosting} and is accessed via a web browser. When the \textit{Widget Adaptor} populates the visualization, it requests data from the \textit{API Gateway}. Furthermore, the updated model will be saved in the database via the \textit{API Gateway} when the user customizes the dashboard design.

The backend includes the server-side of the application and the supporting database.
The core of the backend is the \textit{API Gateway}, which handles requests from the frontend, saves new customization, and returns the requested data. The data is stored in a MongoDB database, and two \textit{Data Collections} are designed to store the Mod2Dash dashboard model and the corresponding random metric values.

The frontend web hosting and backend service run on the same cloud server. 

\subsection{Technologies} \label{subsec:technologies}
The implementation of the frontend is based on ReactJS\footnote{ReactJS: https://reactjs.org/} (a JavaScript framework), the main UI frameworks include Ant Design\footnote{Ant Design: https://ant.design/} (a UI framework with chart libraries), React Grid Layout\footnote{React Grid Layout: https://github.com/react-grid-layout/react-grid-layout} (a grid layout control system for React), React Gauge Chart\footnote{React Gauge Chart: https://github.com/Martin36/react-gauge-chart} (a library for Gauge Chart) and nivo\footnote{nivo: https://nivo.rocks/} (a rich set of dataviz components), the request sent from frontend to backend is handled by axios\footnote{axios: https://github.com/axios/axios} (a promise based HTTP client for the browser and node.js).

All services in the backend are disclosed by RESTful APIs, which are written by Flask\footnote{Flask: https://palletsprojects.com/p/flask/} (a lightweight web application framework) and Swagger\footnote{Swagger: https://swagger.io/} (an interface description language for describing RESTful APIs). The backend service is deployed by Gunicorn\footnote{Gunicorn: https://gunicorn.org/} (a Python WSGI HTTP Server). We use MongoDB\footnote{MongoDB: https://www.mongodb.com/} (a NoSQL database program) to store all the data and configuration. 

The whole system can be deployed on an Amazon Elastic Compute Cloud\footnote{Amazon EC2: https://aws.amazon.com/ec2/}.

\section{Evaluation} \label{sec:evaluation}
This section discusses the evaluation methodology and the results. This evaluation aims to quantify the effectiveness of Mod2Dash in real-world dashboard generation scenarios, namely, to answer the research question \textit{"How effective that Mod2Dash can represent and reproduce the dashboard design decisions?"}.

\subsection{Methodology}

\subsubsection{Case Study}

Case studies formed the basis of our overall research strategy since case studies are empirical methods for investigating contemporary phenomena in the real-life context \cite{AliBabar2007, Runeson2009, Yinbook2003, Wohlin2021}. The Mod2Dash approach is designed for and applicable across various real-world disciplines with uncontrollable variables in the context. Thus, the case study is an appropriate methodology to perform this evaluation. 
The evaluation was conducted by following the guidelines and checklists for the software engineering case study provided by Runeson and Höst \cite{Runeson2009}.

To evaluate the effectiveness of Mod2Dash, we conducted a case study on cyber security dashboards for cyber situational awareness (CSA) and decision support in real-world environments. 
We collected and selected 31 cyber security dashboards from the wild, then used our approach to model and rebuild all of these 31 dashboards.
The effectiveness in this context is the ability of Mod2Dash to represent and reproduce diverse real-world dashboard design decisions.

\subsubsection{Real-life Context}

\begin{figure}[!h]
\minipage[h]{0.78\textwidth}
\centering
\includegraphics[width=1\textwidth]{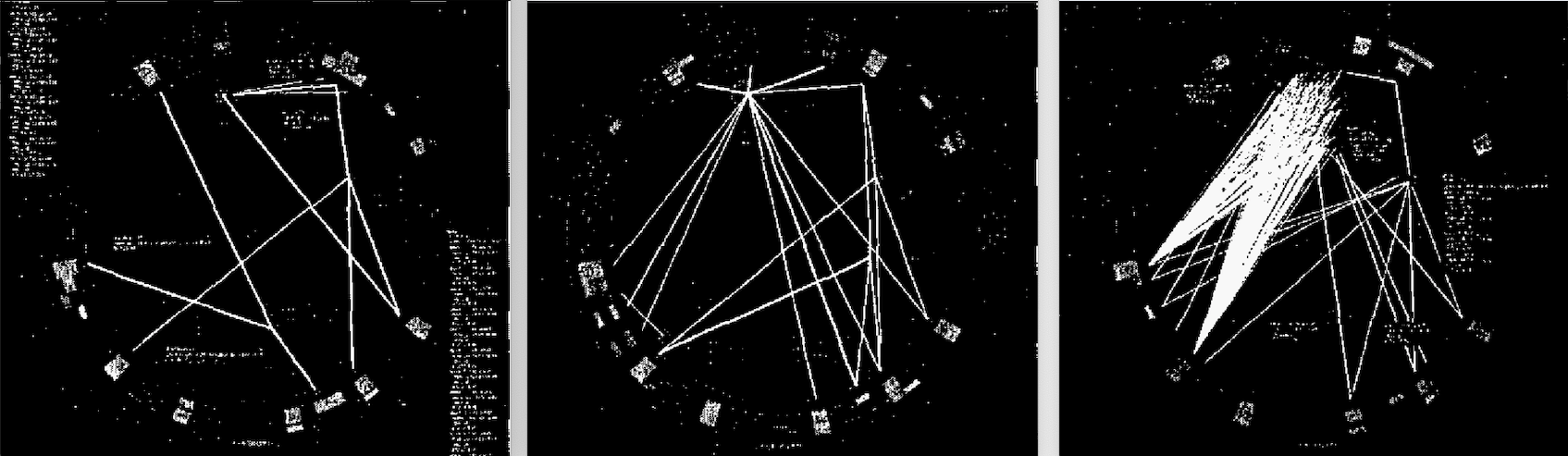}
\caption{CSA visualization that proposed by Livnat et al.\cite{Livnat200592} in 2005.}
\label{2005CSAV}
\endminipage\hfill
\end{figure}

Dashboards are used to monitor and visualize cyber incidents, trends, and threats in the cyber security domain. 
For example, academic researchers propose novel approaches and mechanisms for enhancing CSA, and dashboard prototypes are designed and developed to illustrate their proposals.
Companies in the industry use the dashboards provided by software vendors (e.g., SIEM\footnote{SIEM: Security information and event management}, IDS\footnote{IDS: Intrusion detection system}) to monitor cyber incidents and visualize the key metrics. 
Design professionals collaborate with security analysts to apply new graphic design elements to conceptualize better cyber security dashboards.

\begin{figure}[!htb]
\minipage[t]{0.52\textwidth}
\centering
  \includegraphics[width=1\textwidth]{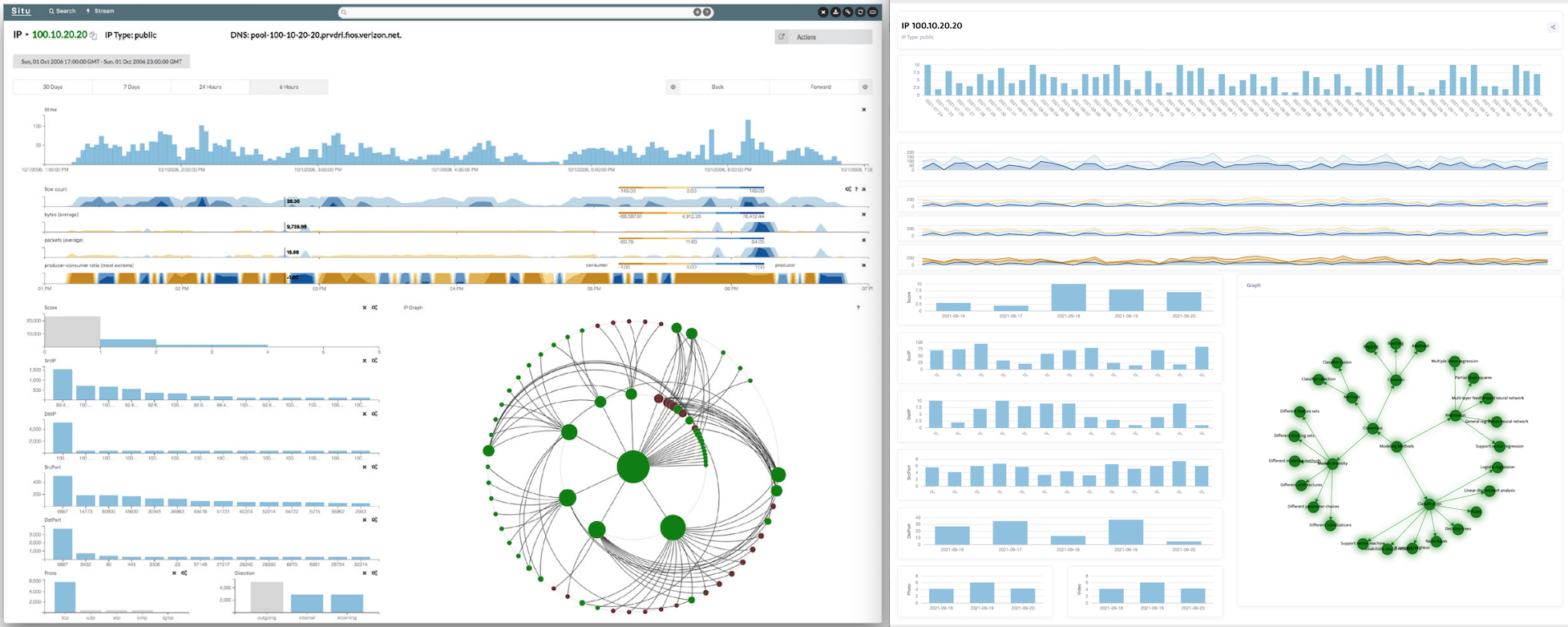}
  \caption{Dashboard that proposed by Goodall et al. \cite{Goodall2019204} in 2019 (left) and the replica generated by Mod2Dash approach (right).}
  \label{situ}

\endminipage\hfill
\minipage[t]{0.46\textwidth}
\centering
  \includegraphics[width=1\textwidth]{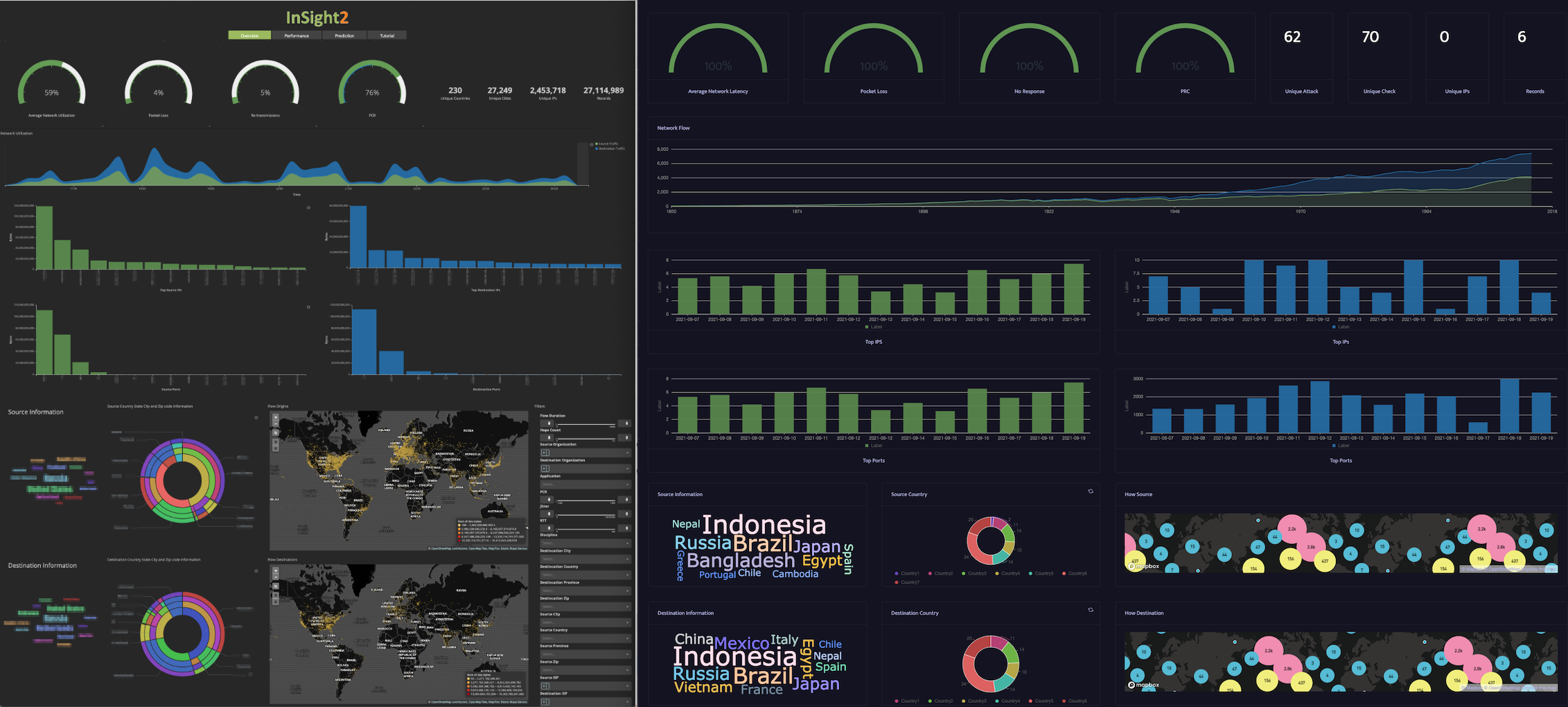}
  \caption{Dashboard that proposed by Kodituwakku, Keller and Gregor \cite{Kodituwakku20201} in 2020 (left) and the replica generated by Mod2Dash approach (right).}
  \label{insight2dash}
\endminipage\hfill

\end{figure}

Utilizing visualization techniques to enhance CSA is a continuously improving process, contributed by cyber data processing and graphical techniques. 
In 2005, Livnat et al. \cite{Livnat200592} presented a concise and scalable representation of the alert attributes in a network intrusion detection system to enhance users' situational awareness (Figure \ref{2005CSAV}). Since 2013, CSA visualization research has been emerging \cite{Jiang2021}. Goodall et al. \cite{Goodall2019204} propose a system to detect and visualize attacks and abnormal network activity. The proposed dashboard (Figure \ref{situ}) enables operators to identify and investigate the situation and provides context to help operators understand the situation. 
Kodituwakku et al. \cite{Kodituwakku20201} propose a visual analysis platform (Figure \ref{insight2dash}) for visualizing real-time data in large-scale networks to provide computer network CSA for security analysts and researchers. 

To the best of our experience and knowledge, Mod2Dash is the first trial to use the model-driven automated approach to build dashboards in the cyber security domain.

\subsection{Data Collection}
\subsubsection{Dashboards Collection Process}
We widely collected cyber-related dashboards as part of the effort to make the dashboards dataset practical, comprehensive, and reflect real-life context. Each author explored and downloaded his own set of cyber security visualizations that he believed might qualify as dashboards. We then combined these dashboards with the dashboard collection from \cite{Sarikaya2019} and used inclusion and exclusion criteria (Table~\ref{tab:inclusionExclusion} in Appendix) to identify whether a dashboard could be selected for the final collection. The final dashboard collection included 31 dashboards after applying the inclusion and exclusion criteria.

\subsubsection{Design Decisions Extraction.} 
Each dashboard captures a set of design decisions, reflecting the dashboard practitioners' knowledge and understanding of their CSA and decision support. In order to simulate the process of creating dashboards based on the practitioners' models, we must first examine the practitioners' design decisions. Therefore, we carried out an analysis of design decisions among these 31 dashboards.

We followed a predefined protocol and procedure to ensure the design decisions we extracted were accurate and precise. First, two authors read through all the dashboards one by one and recorded potential design parameters that could match the Mod2Dash dashboard language. Second, two authors performed the formal extraction process. Third, all authors checked and confirmed that all data was correctly extracted.

\subsubsection{Proof-of-Concept Development}
Based on the Mod2Dash design presented in Section \ref{sec:mod2dash} and the utilization of the techniques presented in Section \ref{sec:implementation}, we developed a proof-of-concept of Mod2Dash.

We implemented the dashboard models in JSON format, which the frontend could directly access and store in the database.
We then developed the automated dashboard generation mechanism based on reading the JSON model from the API and translating it into a web-based application by multiple levels of interpreters.
We also developed GUI-driven customization features to enable users to continue improving the dashboards.

\subsubsection{Dashboards Reconstruct} 
We employed the Mod2Dash approach to reconstruct the selected 31 cyber security dashboards. 
Our goal was to capture as many practitioners' design decisions as possible in the Mod2Dash dashboard model and replicate these dashboards in a web-based application as closely as possible to the original dashboards.
We followed the steps below to build and improve a dashboard. Figure \ref{workflow} also illustrates the overview workflow for building a dashboard by using the Mod2Dash approach.

\begin{itemize}
    \item The authors examined the original dashboard and captured the design decisions in the model.
    \item The authors stored the model in the database.
    \item The frontend read from the backend and built the dashboard according to the model.
    \item The authors checked the initial dashboard implementation result and compared it with the original dashboard.
    \item The authors kept improving the dashboard via the GUI until the dashboard presents as similar as possible to the baseline dashboard.
\end{itemize}

\begin{table}[b] \caption{\label{tab:evaluation}Evaluation form used for each dashboard.}
\begin{tabular}{p{0.35\linewidth} | p{0.65\linewidth}}
\hline
\multicolumn{2}{c}{Compare (Number)}                                                                                                                                                                                                                                                  \\ \hline
\multicolumn{1}{c|}{Major Design Decisions}         & The participants were required to count what they think are the major design decisions on original dashboards and replica dashboards. Major design decisions include data representation format, and visualization techniques. \\ \hline
\multicolumn{1}{c|}{Minor Design Decisions}         & The participants were required to count what they think are the minor design decisions on original dashboards and replica dashboards. Minor design decisions include dashboard layout, color selection, and theme.             \\ \hline
\multicolumn{1}{c|}{Interactions}                   & The participants were required to count what they think are the interactions, such as buttons, and icons.                                                                                                                      \\ \hline
\multicolumn{2}{c}{Impression (Rate 0-10)}                                                                                                                                                                                                                                              \\ \hline
\multicolumn{1}{c|}{Overall Performance Impression} & Overall impression about how well Mod2Dash rebuilds the dashboard.                                                                                                                                                              \\ \hline
\multicolumn{1}{c|}{Correctness}                    & The correctness when Mod2Dash rebuilds the dashboard.                                                                                                                                                                           \\ \hline
\multicolumn{1}{c|}{Visualization}                  & The visualization rating when comparing replica dashboard with the original one.                                                                                                                                                \\ \hline
\multicolumn{1}{c|}{Understandability}              & The rating about whether the visualization is understandable when comparing Mod2Dash with the original one.                                                                                                                     \\ \hline
\multicolumn{1}{c|}{Expressiveness}                 & The rating about whether the visualization is expressive when comparing Mod2Dash with the original one.                                                                                                                         \\ \hline
\multicolumn{2}{c}{Feedback}                                                                                                                                                                                                                                                          \\ \hline
\multicolumn{2}{ p{1\linewidth}}{General feedback about how well Mod2Dash rebuilds this dashboard, detailed feedback about how Mod2Dash can be improved.}                                                                                                                                           \\ \hline
\end{tabular}
\end{table}

\subsubsection{Human-assisted Study}

End users consume the dashboard design, and the performance of the proposed approach in the cyber context should be examined by real users in the cyber security and data analysis domain. 
Thus, we conducted a human-assisted study to quantify the effectiveness of the Mod2Dash approach. Furthermore, this study aims to identify matching dashboard design decisions in the original baseline dashboard and the dashboard constructed by Mod2Dash.

We recruited more than one participant to reduce errors in identifying the design decisions.
Four Ph.D. students were recruited to conduct this evaluation. They have an average of 2.75 years of experience in the data analysis and cyber security domains.

The baseline dashboards were collected from the wild in image format, which means that the interactive features may not be interactive. We relied on the dashboard image and our understanding to analyze interactive features during the data extraction process. 
Equivalently, in the evaluation, 
we screenshotted the dashboards and printed out both original and replica dashboards on the same page for quick and easy comparison. 
We also provided a form (Table~\ref{tab:evaluation}) with a set of metrics for each dashboard to help the participants follow the procedure.

The evaluation session began with a 10-minute introduction of visualization background knowledge, after which the evaluation protocol and rules were briefly discussed. 
Then, the participants were required to compare a pair of dashboards one by one, decompose the elements in the dashboards in their minds, and find the major design decisions (e.g., data representation format and visualization techniques), minor design decisions (e.g., dashboard layout, color selection, and theme), and interactions on the pair dashboards. 
The participants counted the design decisions in the original baseline dashboard first and then looked for matching design decisions in the replica dashboard.
Participants may have different understandings of the design, and they were encouraged to identify the items based on their knowledge and understanding. However, the design decisions found in the replica dashboards are not allowed to be more than the original baseline dashboard. 
The participants were requested to fill in the \textit{Compare} section (Table~\ref{tab:evaluation}) during the counting and rate the performance impression about the current dashboard reconstruction (\textit{Impression} section in Table~\ref{tab:evaluation}) after reviewing each pair of dashboards. The rating items include overall performance impression, correctness, visualization, understandability, expressiveness. The impression ratings should be in the range of 0 to 10. 
The participants were also required to give feedback (\textit{Feedback} section in Table~\ref{tab:evaluation}) for each dashboard reconstruction. 
Each evaluation session for a participant took about 2 hours. We awarded a \$60 gift card to each participant.

\subsection{Results}

This subsection discusses the results of the real-world dashboard collection, the reconstruction of 31 cyber security dashboards, and the effectiveness of Mod2Dash.

\subsubsection{Demographics} \label{subsec:demographics}
The cyber security dashboards (contemporary phenomena in the real-life context of this case study) are coming from diverse sources, and the characteristics of these dashboards are also diverse and complex.

The sources of cyber security dashboards are diverse. As shown in Table~\ref{tab:distribution}, we categorized the source distribution of these dashboards: (1) research papers with the topic of cyber security visualization or dashboard (5 dashboards, 16.13\%), (2) industry practices in cyber security solution system (20 dashboards, 64.52\%), with characters of "SIEM", "Security Dashboard" (e.g., SolarWinds\footnote{SolarWinds: https://www.solarwinds.com/}, OptimEyes.ai\footnote{OptimEyes.ai: https://optimeyes.ai/}, and IBM QRadar\footnote{IBM QRadar: https://www.ibm.com/au-en/security/security-intelligence/qradar}), (3) Security dashboard visual design in the design sharing platform Dribbble\footnote{Dribbble: https://dribbble.com/} (4 dashboards, 12.90\%), (4) Cyber reporting Excel template (2 dashboards, 6.45\%).

\begin{table}[h] \caption{\label{tab:distribution}Cyber security dashboard collected from different resources.}
\begin{tabular}{lllp{8cm}}
\hline
Source    & Number & Percentage & Description                                                                \\ \hline
Academia  & 5      & 16.13\%    & Research papers with the topic of cyber security visualization or dashboard. \\ \hline
Industry  & 20     & 64.52\%    & Industry practices in cyber security solution products     .                   \\ \hline
Design    & 4      & 12.90\%    & Cyber dashboard visual designs in the design sharing platform "Dribbble".                           \\ \hline
Reporting & 2      & 6.45\%     & Excel reporting templates for cyber security.                                                   \\ \hline
\end{tabular}
\end{table}

We further examined and analyzed these 31 cyber security dashboards to understand the practitioners' design decisions when they designed and built the dashboards. We found that the characteristics of these dashboards are also diverse and complex.

\begin{itemize}

    \item \textbf{Layout} is how the components are located in a multiple-view visualization (MV), it is critical design decision for dashboards. Chen et al. \cite{Chen2021} identified 98 unique layouts from 360 MV designs. Based on their findings, we identified and compressed the dashboard layouts for these 31 dashboards in 4 categories: \textit{Standard Grid} \includegraphics[height=2ex]{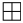} (3 dashboards, 9.68\%) is the widgets in the dashboard are aligned both horizontally and vertically. \textit{Row Oriented} \includegraphics[height=2ex]{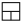} means the dashboard widgets are populated row by row (horizontal), widgets could be not aligned vertically, and \textit{Row Oriented} \includegraphics[height=2ex]{image/l3.png} (18 dashboards, 58.06\%) is the most popular design layout for cyber security dashboards. \textit{Column Oriented} \includegraphics[height=2ex]{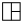} (4 dashboards, 12.90\%) means the dashboard widgets are populated column by column (vertial), widgets could be not aligned horizontally. \textit{Disordered} \includegraphics[height=2ex]{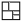} (6 dashboard, 19.35\%) layouts do not strictly follow horizontal or vertical design.
    
    \item \textbf{Theme} is a set of color combinations for a dashboard. The default setting in the visualization development is to display black text on a white background\cite{darkmode}. The dark mode became a UI design trend after the release of the macOS Mojave update \cite{darkmode2, darkmode3}. We found that 22 of 31 cyber security dashboards (70.97\%) use a light theme, and 9 dashboards (29.03\%) use a dark theme. 

    \item \textbf{Visualization Techniques} are how the data is visualized, and they are the lowest level elements inside a widget. We identified 18 different Visualization Techniques (i.e., Single Value, Table, Gauge, Area Chart, Column Chart, Word Cloud, Ring, Map, Composite Chart, Scatter, Radial Tree, Pie Chart, Bar Chart, Treemap, Line Chart, Bullet, Sankey, and Radar) during the analysis. We found that \textit{Table}, \textit{Column Chart}, and \textit{Single Value} are the most popular techniques used in these dashboards. The detailed distribution of these techniques used in the collection can be found in Figure~\ref{vistech} in the Appendix.
    
    \item \textbf{Widget Number} and \textbf{Visualization Technique Number} in a dashboard indicate the data complexity and the metrics that are monitored in a dashboard. We considered a widget as a basic unit of data fact, and we treated different data representation formats (e.g., text, table) and different chart types as different \textit{Visualization Technique}. As a result, we found that the maximum number of widgets in a dashboard is 19, and the minimal number is 3, with an average of 8 widgets among these dashboards. Meanwhile, a dashboard is a comprehensive use of different visualization techniques. We found the maximal number using different \textit{Visualization Technique} in a single dashboard is 7, and the minimal is 2, with an average of 4. For example, the dashboard in \cite{Kodituwakku20201}  (Figure \ref{insight2dash}) uses 19 widgets to monitor real-time and large-scale networks by applying 7 different visualization techniques (i.e., Gauge, Single Value, Area Chart, Column Chart, Word Cloud, Ring, and Map).
    
    \item \textbf{Interactions} in this study are focused more on dashboard-level and widget-level rather than the interactions in a chart. We extracted interactions from the dashboard description and icons in the dashboard, we found 8 different interaction types: \textit{filter}, \textit{zoom}, \textit{share}, \textit{customization}, \textit{detail on demand}, \textit{refresh}, \textit{print}, and \textit{navigation}. We found that 19 dashboards (61.29\%) provide dashboard-level interactions and 16 dashboards (51.61\%) provide widget-level interactions.
    
\end{itemize}

Through the dashboard analysis and design decisions extraction process, we gain insights into the dashboard practitioners' design and their understanding of the decision support and CSA. Furthermore, we have a thorough understanding of dashboards and the diversity of dashboard designs in real-world environments. The detailed data extraction results are illustrated in Table~\ref{tab:dataextration} in the Appendix.

\subsubsection{Dashboard Reconstruction}

A proof-of-concept of the proposed Mod2Dash framework was developed during the case study, and all 31 cyber security dashboards were built based on the framework. 
The reconstruction results can be found in the Figure \ref{situ}, Figure \ref{insight2dash}, Figure \ref{reconstruct}, and Figure \ref{alldashboards} (in the Appendix). 
These pairs of comparison screenshots show that the dashboards built by the Mod2Dash approach represent the designs in the original dashboard well in terms of the visualization techniques, layout, and color combination.

The following subsection will introduce the quantified effectiveness of the Mod2Dash approach in representing and reproducing these real-world dashboards.

\begin{figure}[h]
\begin{minipage}[h]{0.98\textwidth}
\centering
\includegraphics[width=1\textwidth]{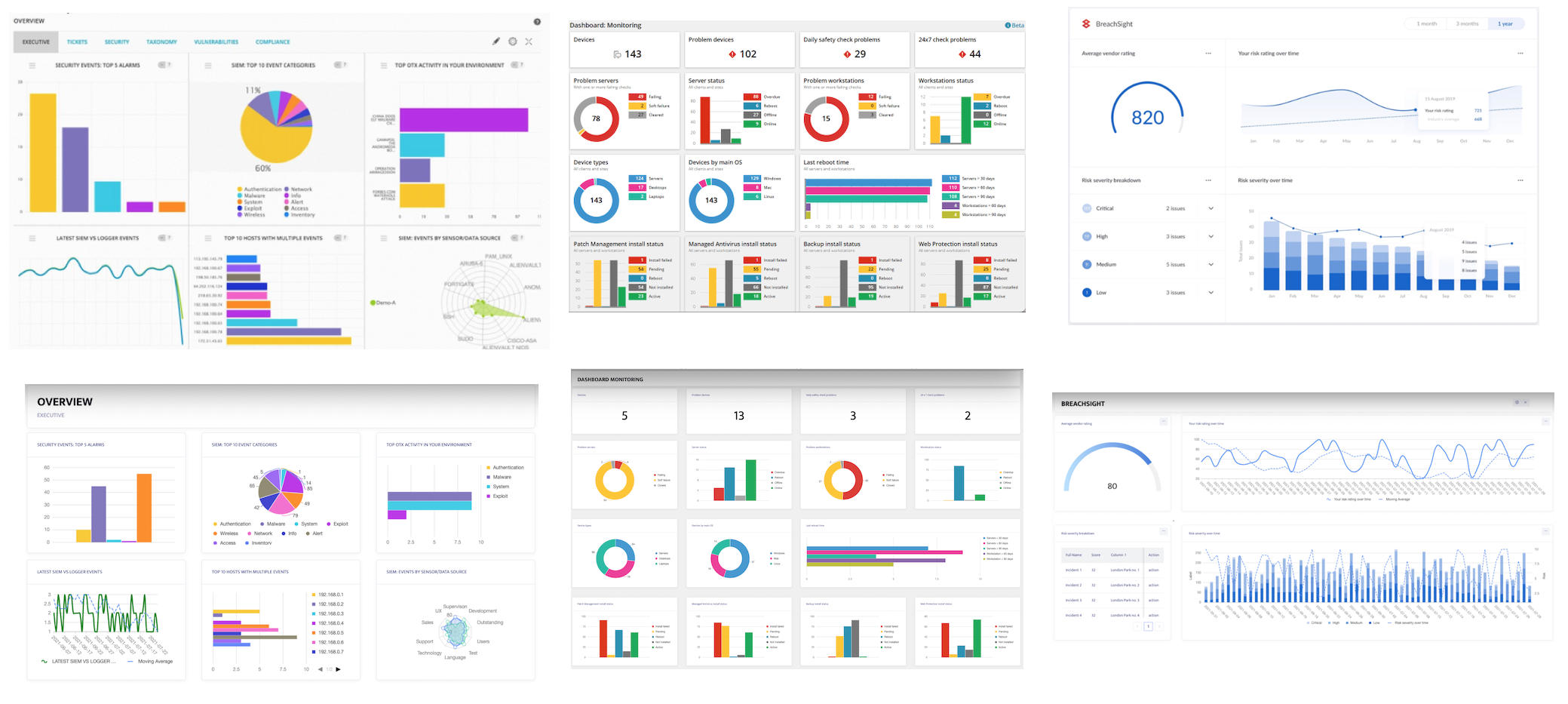}
\caption{Comparison screenshots of the original dashboards and replicas. The first row is the original baseline dashboards, and the second row is the dashboards generated by the Mod2Dash approach.}
\label{reconstruct}
\end{minipage}\hfill%
\end{figure}

\subsubsection{Reproduction Rate}

The reproduction rate is a metric to measure the effectiveness of Mod2Dash. It is calculated based on the matching design decision numbers that participants identified during the human-assisted study session. It illustrates how effective the Mod2Dash approach is adapted to diverse real-world dashboard environments regarding dashboard model representation for design decisions and the visual dashboard components generated based on the dashboard model.

\begin{figure}[t]
\begin{minipage}[t]{0.98\textwidth}
\centering
\includegraphics[width=1\textwidth]{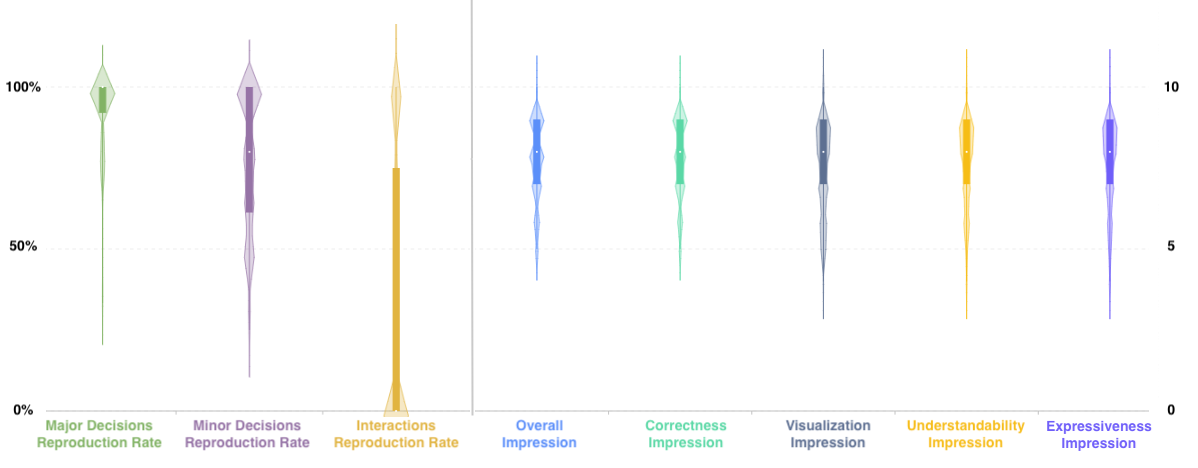}
\caption{The reproduction rate for major design decisions, minor design decisions, interactions, and the impression ratings.}
\label{boxplot}
\end{minipage}\hfill%
\end{figure}

The participants found 843 major design decisions in the original dashboards and 802 in replicas, with a replication rate of 95.13\%. The participants identified a total of 803 and 617 minor design decisions in the original dashboards and the replica dashboards, rendering a reproduction rate of 76.84\%. Participants found 189 interaction buttons or icons in the original dashboards and 85 corresponding replicas interactions, making the overall reproduction rate 44.97\%. The raw data from the human-assisted study composes the Figure ~\ref{boxplot}.

The high reproduction rates demonstrate the effectiveness of the Mod2Dash dashboard model in capturing real-world dashboard design decisions. Furthermore, it also shows that Mod2Dash can effectively generate real-world dashboards that accurately represent major and minor design decisions with an acceptable interaction representation.

\subsubsection{Performance Rating} 
Participants were requested to rate their impression of the Mod2Dash approach. The impression rating was evaluated from overall performance impression, correctness, visualization, understandability, and expressiveness. 

The maximal rating that the participants gave for these metrics was 10, and the minimal rating was 4, with a median score of 8. 
Despite a few poor ratings indicating that few Mod2Dash-produced dashboards perform not as well as the baselines, the overall impression is good. Figure \ref{boxplot} shows the impression ratings, which is generated from the raw data.

The ratings from the participants also show that Mod2Dash performs well in presenting real-world dashboards.

\subsubsection{Feedback} 
Participants also provided qualitative feedback based on their observations. After further analyzing their comments, we gained insights into improving the mechanism in the future.

\textbf{Pros:} During the evaluation, two of the participants commented that some of the replicas are superior to the original baseline dashboards in terms of visualization and expressiveness. 
In the evaluation forms, we could find words such as \textit{"looks good to me", "this replication captured almost perfectly"} to praise the Mod2Dash approach.

\textbf{Cons:} 
There are also some drawbacks that the participants had identified. All the participants pointed out that the color scheme needs to be improved, and the font size is not visible well in the replicas. Mod2Dash provides configuration for the color in the charts. The model can describe all the colors in a chart (see the comparison in Figure ~\ref{reconstruct}), but the model fails in dealing with other hard-coded color settings in the dashboard. For example, the original dashboard uses different colors for different rows in a PowerPoint template. Although Mod2Dash provides the font size configuration for \textit{Single Value} widget, our current model is still unable to capture and configure other font sizes such as font size of the title in a table, font size of value number in a \textit{Ring Chart}, or font size of the legend.

Participants also raised the issue of missing indicators. For example, there was a widget using a condensed form of visualization, combining \textit{Line Chart}, \textit{Column Chart}, and \textit{Area Chart} in one single widget. Our current technology is not able to recur this widget. Furthermore, some dashboards are hard-coded with different colors for different values in a \textit{Table} to denote the importance of the information, but Mod2Dash cannot capture this indicator as well.

\subsection{Threat to Validity}

This section discusses the limitations and main threats to the internal and external validity of this study. 
The case study was carefully designed and carried out to prevent undesired factors from affecting the outcome. However, a few essential details need to be considered when replicating the experiment.

\begin{itemize}

    \item Although we argue that we collect and select dashboards widely from different sources, researchers' subjective judgment may influence the dashboard selection. To reduce the influence of this issue, we relied on pre-defined inclusion and exclusion criteria (Table~\ref{tab:inclusionExclusion} in the Appendix) when collecting and selecting dashboards.
    \item In the study, we targeted the gaps in capturing the models or design decisions in the heads of dashboard practitioners and automatically generating the dashboard based on the model. It primarily focused on dashboard visualization, and the data used in this study is mock data or created by a random algorithm. Therefore, the results may vary regarding the chart color or shape if different data models are used in the replicating experiments.
    \item The judgments of an interaction icon or a component in the dashboard belongs to which interaction category are also subjective when extracting design decisions from the dashboards, especially we have to use the dashboard screenshots to analyze the interactions. For example, an icon of three dots could mean \textit{detail on demand} or \textit{configuration} interaction feature. To address this problem, we designed a process for data extraction that extracts data in multiple steps and with multiple authors' consistent input.
    \item Another limitation that needs to be discussed is the design of the human-assisted study. We relied on the participants' design decision counting to evaluate the effectiveness of Mod2Dash. 
    The printed version of pair dashboards and the participant's experience and knowledge may influence the judgments regarding the design decision of a particular component.
    Consequently, we recruited four participants to count the number of design decision matching to reduce this issue's influence. The plan and development of the interview guides and selection of participants can be improved in further formal evaluation.

\end{itemize}

\section{Conclusions} \label{sec:conclusions}

This paper proposes Mod2Dash, a framework for model-driven dashboard generation. Researchers and developers can use it to quickly model, prototype, and validate dashboard designs.
The Mod2Dash approach contains three main components.
First, the dashboard visualization language helps dashboard practitioners capture their implicit design decisions. Second, the automated dashboard generation mechanism builds the operational dashboards, which is helpful for dashboard practitioners to communicate their design. Third, a GUI-driven customization approach helps users continuously improve the dashboard design.
We presented a fully functional framework and evaluated it in a case study. 
The evaluation result shows that Mod2Dash can effectively capture the real-world dashboard design decisions and generate real-world dashboards.

As a practical and comprehensive means of presenting critical information, we believe that dashboard is an important and challenging research topic.
One of the interesting future directions that can be based on the outcome of this paper is dashboard visualization recommendation. First, a rule-based or machine learning-based recommendation approach can be derived by analyzing or learning from more real-world dashboard designs. Then the recommendation approach suggests dashboard models by following the Mod2Dash dashboard language based on the practitioners' input. Finally, the dashboards are automatically generated by Mod2Dash so that the recommended designs can be quickly evaluated.

\begin{acks}
The work has been supported by the Cyber Security Research Centre Limited whose activities are partially funded by the Australian Government’s Cooperative Research Centres Programme.
\end{acks}

\bibliographystyle{ACM-Reference-Format}
\bibliography{sample-base}


\begin{thebibliography}{44}


\ifx \showCODEN    \undefined \def \showCODEN     #1{\unskip}     \fi
\ifx \showDOI      \undefined \def \showDOI       #1{#1}\fi
\ifx \showISBNx    \undefined \def \showISBNx     #1{\unskip}     \fi
\ifx \showISBNxiii \undefined \def \showISBNxiii  #1{\unskip}     \fi
\ifx \showISSN     \undefined \def \showISSN      #1{\unskip}     \fi
\ifx \showLCCN     \undefined \def \showLCCN      #1{\unskip}     \fi
\ifx \shownote     \undefined \def \shownote      #1{#1}          \fi
\ifx \showarticletitle \undefined \def \showarticletitle #1{#1}   \fi
\ifx \showURL      \undefined \def \showURL       {\relax}        \fi
\providecommand\bibfield[2]{#2}
\providecommand\bibinfo[2]{#2}
\providecommand\natexlab[1]{#1}
\providecommand\showeprint[2][]{arXiv:#2}

\bibitem[\protect\citeauthoryear{{Ali Babar}, Verner, and Nguyen}{{Ali Babar}
  et~al\mbox{.}}{2007}]%
        {AliBabar2007}
\bibfield{author}{\bibinfo{person}{Muhammad {Ali Babar}},
  \bibinfo{person}{June~M. Verner}, {and} \bibinfo{person}{Phong~Thanh
  Nguyen}.} \bibinfo{year}{2007}\natexlab{}.
\newblock \showarticletitle{{Establishing and maintaining trust in software
  outsourcing relationships: An empirical investigation}}.
\newblock \bibinfo{journal}{\emph{Journal of Systems and Software}}
  \bibinfo{volume}{80}, \bibinfo{number}{9} (\bibinfo{year}{2007}),
  \bibinfo{pages}{1438--1449}.
\newblock
\showISSN{01641212}
\urldef\tempurl%
\url{https://doi.org/10.1016/j.jss.2006.10.038}
\showDOI{\tempurl}


\bibitem[\protect\citeauthoryear{Alzoubi, Kelley, Baran, Gilbert, {Karabulut
  Ilgu}, and Jiang}{Alzoubi et~al\mbox{.}}{2021}]%
        {Alzoubi2021}
\bibfield{author}{\bibinfo{person}{Dana Alzoubi}, \bibinfo{person}{Jameel
  Kelley}, \bibinfo{person}{Evrim Baran}, \bibinfo{person}{Stephen~B. Gilbert},
  \bibinfo{person}{Aliye {Karabulut Ilgu}}, {and} \bibinfo{person}{Shan
  Jiang}.} \bibinfo{year}{2021}\natexlab{}.
\newblock \showarticletitle{{TeachActive Feedback Dashboard: Using Automated
  Classroom Analytics to Visualize Pedagogical Strategies at a Glance}}.
\newblock \bibinfo{journal}{\emph{Conference on Human Factors in Computing
  Systems - Proceedings}} (\bibinfo{year}{2021}).
\newblock
\showISBNx{9781450380959}
\urldef\tempurl%
\url{https://doi.org/10.1145/3411763.3451709}
\showDOI{\tempurl}


\bibitem[\protect\citeauthoryear{Berman}{Berman}{2018}]%
        {kibanatutorial}
\bibfield{author}{\bibinfo{person}{Daniel Berman}.}
  \bibinfo{year}{2018}\natexlab{}.
\newblock \bibinfo{booktitle}{\emph{A Kibana Tutorial – Part 2: Creating
  Visualizations}}.
\newblock
\urldef\tempurl%
\url{https://logz.io/blog/kibana-tutorial-2/#kibanaaggregations}
\showURL{%
Retrieved Aus 31, 2021 from \tempurl}


\bibitem[\protect\citeauthoryear{Chen, Zeng, Lin, Ai-Maneea, Roberts, and
  Chang}{Chen et~al\mbox{.}}{2021}]%
        {Chen2021}
\bibfield{author}{\bibinfo{person}{Xi Chen}, \bibinfo{person}{Wei Zeng},
  \bibinfo{person}{Yanna Lin}, \bibinfo{person}{Hayder~Mahdi Ai-Maneea},
  \bibinfo{person}{Jonathan Roberts}, {and} \bibinfo{person}{Remco Chang}.}
  \bibinfo{year}{2021}\natexlab{}.
\newblock \showarticletitle{{Composition and configuration patterns in
  multiple-view visualizations}}.
\newblock \bibinfo{journal}{\emph{IEEE Transactions on Visualization and
  Computer Graphics}} \bibinfo{volume}{27}, \bibinfo{number}{2}
  (\bibinfo{year}{2021}), \bibinfo{pages}{1514--1524}.
\newblock
\showISSN{19410506}
\urldef\tempurl%
\url{https://doi.org/10.1109/TVCG.2020.3030338}
\showDOI{\tempurl}
\showeprint[arxiv]{2007.15407}


\bibitem[\protect\citeauthoryear{Chowdhary, Palpanas, Pinel, Chen, and
  Wu}{Chowdhary et~al\mbox{.}}{2006}]%
        {Chowdhary2006}
\bibfield{author}{\bibinfo{person}{Pawan Chowdhary}, \bibinfo{person}{Themis
  Palpanas}, \bibinfo{person}{Florian Pinel}, \bibinfo{person}{Shyh~Kwei Chen},
  {and} \bibinfo{person}{Frederick~Y. Wu}.} \bibinfo{year}{2006}\natexlab{}.
\newblock \showarticletitle{{Model-driven dashboards for business performance
  reporting}}.
\newblock \bibinfo{journal}{\emph{Proceedings - IEEE International Enterprise
  Distributed Object Computing Workshop, EDOC}} \bibinfo{number}{November}
  (\bibinfo{year}{2006}), \bibinfo{pages}{374--383}.
\newblock
\showISBNx{076952558X}
\showISSN{15417719}
\urldef\tempurl%
\url{https://doi.org/10.1109/EDOC.2006.34}
\showDOI{\tempurl}


\bibitem[\protect\citeauthoryear{{De Croon}, Leeuwenberg, Aerts, Moens, {Vanden
  Abeele}, and Verbert}{{De Croon} et~al\mbox{.}}{2021}]%
        {DeCroon2021}
\bibfield{author}{\bibinfo{person}{Robin~De {De Croon}},
  \bibinfo{person}{Artuur Leeuwenberg}, \bibinfo{person}{Jan Aerts},
  \bibinfo{person}{Marie~Francine Moens}, \bibinfo{person}{Vero~Vanden {Vanden
  Abeele}}, {and} \bibinfo{person}{Katrien Verbert}.}
  \bibinfo{year}{2021}\natexlab{}.
\newblock \showarticletitle{{TIEVis: A visual analytics dashboard for temporal
  information extracted from clinical reports}}.
\newblock \bibinfo{journal}{\emph{International Conference on Intelligent User
  Interfaces, Proceedings IUI}} (\bibinfo{year}{2021}),
  \bibinfo{pages}{34--36}.
\newblock
\showISBNx{9781450380188}
\urldef\tempurl%
\url{https://doi.org/10.1145/3397482.3450731}
\showDOI{\tempurl}


\bibitem[\protect\citeauthoryear{Delis and Roussopoulos}{Delis and
  Roussopoulos}{1992}]%
        {Delis}
\bibfield{author}{\bibinfo{person}{Alexios Delis} {and} \bibinfo{person}{Nick
  Roussopoulos}.} \bibinfo{year}{1992}\natexlab{}.
\newblock \showarticletitle{{Performance and Scalability of Client Server
  Database Architectures}}.
\newblock \bibinfo{journal}{\emph{Proceedings of the 18th VLDB Conference}}
  (\bibinfo{year}{1992}).
\newblock


\bibitem[\protect\citeauthoryear{Dibia and Demiralp}{Dibia and
  Demiralp}{2018}]%
        {Dibia2018a}
\bibfield{author}{\bibinfo{person}{Victor Dibia} {and}
  \bibinfo{person}{{\c{C}}aǧatay Demiralp}.} \bibinfo{year}{2018}\natexlab{}.
\newblock \showarticletitle{{Data2Vis: Automatic generation of data
  visualizations using sequence to sequence recurrent neural networks}}.
\newblock \bibinfo{journal}{\emph{arXiv}} \bibinfo{number}{June 2019}
  (\bibinfo{year}{2018}), \bibinfo{pages}{33--46}.
\newblock
\showISSN{23318422}


\bibitem[\protect\citeauthoryear{Few}{Few}{2004}]%
        {Few2004}
\bibfield{author}{\bibinfo{person}{Stephen. Few}.}
  \bibinfo{year}{2004}\natexlab{}.
\newblock \showarticletitle{{Dashboard Confusion}}.
\newblock \bibinfo{journal}{\emph{Perceptual Edge}} (\bibinfo{year}{2004}),
  \bibinfo{pages}{1--4}.
\newblock
\showISSN{15243621}
\urldef\tempurl%
\url{https://www.perceptualedge.com/articles/ie/dashboard_confusion.pdf%0Ahttp://72.251.211.178/articles/visual_business_intelligence/dboard_confusion_revisited.pdf}
\showURL{%
\tempurl}


\bibitem[\protect\citeauthoryear{Few}{Few}{2013}]%
        {Stephen}
\bibfield{author}{\bibinfo{person}{Stephen Few}.}
  \bibinfo{year}{2013}\natexlab{}.
\newblock \bibinfo{booktitle}{\emph{Information Dashboard Design}
  (\bibinfo{edition}{second edition} ed.)}.
\newblock \bibinfo{publisher}{Analytics Press}, \bibinfo{address}{PO Box 4933,
  EI Dorado Hills, CA 95762}.
\newblock


\bibitem[\protect\citeauthoryear{Goodall, Ragan, Steed, Reed, Richardson,
  Huffer, Bridges, and Laska}{Goodall et~al\mbox{.}}{2019}]%
        {Goodall2019204}
\bibfield{author}{\bibinfo{person}{J~R Goodall}, \bibinfo{person}{E~D Ragan},
  \bibinfo{person}{C~A Steed}, \bibinfo{person}{J~W Reed}, \bibinfo{person}{G~D
  Richardson}, \bibinfo{person}{K~M~T Huffer}, \bibinfo{person}{R~A Bridges},
  {and} \bibinfo{person}{J~A Laska}.} \bibinfo{year}{2019}\natexlab{}.
\newblock \showarticletitle{{Situ: Identifying and explaining suspicious
  behavior in networks}}.
\newblock \bibinfo{journal}{\emph{IEEE Transactions on Visualization and
  Computer Graphics}} \bibinfo{volume}{25}, \bibinfo{number}{1IEEE Transactions
  on Visualization and Computer Graphics} (\bibinfo{year}{2019}),
  \bibinfo{pages}{204--214}.
\newblock
\urldef\tempurl%
\url{https://doi.org/10.1109/TVCG.2018.2865029}
\showDOI{\tempurl}


\bibitem[\protect\citeauthoryear{Jiang, Jayatilaka, Nasim, Grobler, Zahedi, and
  Babar}{Jiang et~al\mbox{.}}{2021}]%
        {Jiang2021}
\bibfield{author}{\bibinfo{person}{Liuyue Jiang}, \bibinfo{person}{Asangi
  Jayatilaka}, \bibinfo{person}{Mehwish Nasim}, \bibinfo{person}{Marthie
  Grobler}, \bibinfo{person}{Mansooreh Zahedi}, {and} \bibinfo{person}{M.~Ali
  Babar}.} \bibinfo{year}{2021}\natexlab{}.
\newblock \showarticletitle{{Systematic Literature Review on Cyber Situational
  Awareness Visualizations}}.
\newblock  (\bibinfo{year}{2021}), \bibinfo{pages}{1--32}.
\newblock
\showeprint[arxiv]{2112.10354}
\urldef\tempurl%
\url{http://arxiv.org/abs/2112.10354}
\showURL{%
\tempurl}


\bibitem[\protect\citeauthoryear{Kintz}{Kintz}{2012}]%
        {Kintz2012}
\bibfield{author}{\bibinfo{person}{Maximilien Kintz}.}
  \bibinfo{year}{2012}\natexlab{}.
\newblock \showarticletitle{{A Semantic Dashboard Description Language for a
  Process-oriented Dashboard Design Methodology}}.
\newblock \bibinfo{journal}{\emph{CEUR Workshop Proceedings}}
  \bibinfo{volume}{947}, \bibinfo{number}{May} (\bibinfo{year}{2012}),
  \bibinfo{pages}{31--36}.
\newblock
\showISSN{16130073}


\bibitem[\protect\citeauthoryear{Kintz, Kochanowski, and Koetter}{Kintz
  et~al\mbox{.}}{2017}]%
        {Kintz2017}
\bibfield{author}{\bibinfo{person}{Maximilien Kintz}, \bibinfo{person}{Monika
  Kochanowski}, {and} \bibinfo{person}{Falko Koetter}.}
  \bibinfo{year}{2017}\natexlab{}.
\newblock \showarticletitle{{Creating user-specific business process monitoring
  dashboards with a model-driven approach}}.
\newblock \bibinfo{journal}{\emph{MODELSWARD 2017 - Proceedings of the 5th
  International Conference on Model-Driven Engineering and Software
  Development}} \bibinfo{volume}{2017-Janua}, \bibinfo{number}{Modelsward}
  (\bibinfo{year}{2017}), \bibinfo{pages}{353--361}.
\newblock
\showISBNx{9789897582103}
\urldef\tempurl%
\url{https://doi.org/10.5220/0006135203530361}
\showDOI{\tempurl}


\bibitem[\protect\citeauthoryear{Kodituwakku, Keller, and Gregor}{Kodituwakku
  et~al\mbox{.}}{2020}]%
        {Kodituwakku20201}
\bibfield{author}{\bibinfo{person}{H.A.D.E. Kodituwakku}, \bibinfo{person}{A
  Keller}, {and} \bibinfo{person}{J Gregor}.} \bibinfo{year}{2020}\natexlab{}.
\newblock \showarticletitle{{Insight2: A modular visual analysis platform for
  network situational awareness in large-scale networks}}.
\newblock \bibinfo{journal}{\emph{Electronics (Switzerland)}}
  \bibinfo{volume}{9}, \bibinfo{number}{10} (\bibinfo{year}{2020}),
  \bibinfo{pages}{1--15}.
\newblock
\urldef\tempurl%
\url{https://doi.org/10.3390/electronics9101747}
\showDOI{\tempurl}


\bibitem[\protect\citeauthoryear{Lab}{Lab}{[n.\,d.]}]%
        {grafanatutorial}
\bibfield{author}{\bibinfo{person}{Grafana Lab}.}
  \bibinfo{year}{[n.\,d.]}\natexlab{}.
\newblock \bibinfo{booktitle}{\emph{Grafana Documentation - List of
  calculations}}.
\newblock
\urldef\tempurl%
\url{https://grafana.com/docs/grafana/latest/panels/calculations-list/}
\showURL{%
Retrieved Aus 31, 2021 from \tempurl}


\bibitem[\protect\citeauthoryear{Livnat, Agutter, Moon, Erbacher, and
  Foresti}{Livnat et~al\mbox{.}}{2005}]%
        {Livnat200592}
\bibfield{author}{\bibinfo{person}{Y Livnat}, \bibinfo{person}{J Agutter},
  \bibinfo{person}{S Moon}, \bibinfo{person}{R~F Erbacher}, {and}
  \bibinfo{person}{S Foresti}.} \bibinfo{year}{2005}\natexlab{}.
\newblock \showarticletitle{{A visualization paradigm for network intrusion
  detection}}. In \bibinfo{booktitle}{\emph{Proceedings from the 6th Annual
  IEEE System, Man and Cybernetics Information Assurance Workshop, SMC 2005}},
  Vol.~\bibinfo{volume}{2005}. \bibinfo{pages}{92--99}.
\newblock
\urldef\tempurl%
\url{https://doi.org/10.1109/IAW.2005.1495939}
\showDOI{\tempurl}


\bibitem[\protect\citeauthoryear{Logz.io}{Logz.io}{2020}]%
        {logzio}
\bibfield{author}{\bibinfo{person}{Logz.io}.} \bibinfo{year}{2020}\natexlab{}.
\newblock \bibinfo{booktitle}{\emph{Grafana vs. Kibana: The Key Differences to
  Know}}.
\newblock
\urldef\tempurl%
\url{https://logz.io/blog/grafana-vs-kibana/}
\showURL{%
Retrieved Aus 31, 2021 from \tempurl}


\bibitem[\protect\citeauthoryear{Michael, Vadim, and Heer}{Michael
  et~al\mbox{.}}{2011}]%
        {Michael2011}
\bibfield{author}{\bibinfo{person}{Bostock Michael},
  \bibinfo{person}{Ogievetsky Vadim}, {and} \bibinfo{person}{Jeffrey Heer}.}
  \bibinfo{year}{2011}\natexlab{}.
\newblock \showarticletitle{{D3: Data-Driven Documents}}.
\newblock  \bibinfo{volume}{2}, \bibinfo{number}{12} (\bibinfo{year}{2011}),
  \bibinfo{pages}{1269--1272}.
\newblock


\bibitem[\protect\citeauthoryear{Palpanas, Chowdhary, Mihaila, and
  Pinel}{Palpanas et~al\mbox{.}}{2007}]%
        {Palpanas2007}
\bibfield{author}{\bibinfo{person}{Themis Palpanas}, \bibinfo{person}{Pawan
  Chowdhary}, \bibinfo{person}{George Mihaila}, {and} \bibinfo{person}{Florian
  Pinel}.} \bibinfo{year}{2007}\natexlab{}.
\newblock \showarticletitle{{Integrated model-driven dashboard development}}.
\newblock \bibinfo{journal}{\emph{Information Systems Frontiers}}
  \bibinfo{volume}{9}, \bibinfo{number}{2-3} (\bibinfo{year}{2007}),
  \bibinfo{pages}{195--208}.
\newblock
\showISSN{13873326}
\urldef\tempurl%
\url{https://doi.org/10.1007/s10796-007-9032-9}
\showDOI{\tempurl}


\bibitem[\protect\citeauthoryear{Poco and Heer}{Poco and Heer}{2017}]%
        {Poco2017}
\bibfield{author}{\bibinfo{person}{Jorge Poco} {and} \bibinfo{person}{Jeffrey
  Heer}.} \bibinfo{year}{2017}\natexlab{}.
\newblock \showarticletitle{{Reverse-Engineering Visualizations: Recovering
  Visual Encodings from Chart Images}}.
\newblock \bibinfo{journal}{\emph{Computer Graphics Forum}}
  \bibinfo{volume}{36}, \bibinfo{number}{3} (\bibinfo{year}{2017}),
  \bibinfo{pages}{353--363}.
\newblock
\showISSN{14678659}
\urldef\tempurl%
\url{https://doi.org/10.1111/cgf.13193}
\showDOI{\tempurl}


\bibitem[\protect\citeauthoryear{Rahman, Adamu, and Harun}{Rahman
  et~al\mbox{.}}{2017}]%
        {Rahman2017}
\bibfield{author}{\bibinfo{person}{Azizah~Abdul Rahman},
  \bibinfo{person}{Yunusa~Bena Adamu}, {and} \bibinfo{person}{Pershella
  Harun}.} \bibinfo{year}{2017}\natexlab{}.
\newblock \showarticletitle{{Review on dashboard application from managerial
  perspective}}.
\newblock \bibinfo{journal}{\emph{International Conference on Research and
  Innovation in Information Systems, ICRIIS}} (\bibinfo{year}{2017}).
\newblock
\showISBNx{9781509030354}
\showISSN{23248157}
\urldef\tempurl%
\url{https://doi.org/10.1109/ICRIIS.2017.8002461}
\showDOI{\tempurl}


\bibitem[\protect\citeauthoryear{Riege, Lee, and Ebrahimi}{Riege
  et~al\mbox{.}}{2019}]%
        {Riege2019}
\bibfield{author}{\bibinfo{person}{Jens Riege}, \bibinfo{person}{Rainier Lee},
  {and} \bibinfo{person}{Nercy Ebrahimi}.} \bibinfo{year}{2019}\natexlab{}.
\newblock \showarticletitle{{Displaying data effectively using an automated
  process dashboard}}.
\newblock \bibinfo{journal}{\emph{IEEE Transactions on Semiconductor
  Manufacturing}} \bibinfo{volume}{32}, \bibinfo{number}{4}
  (\bibinfo{year}{2019}), \bibinfo{pages}{530--537}.
\newblock
\showISSN{15582345}
\urldef\tempurl%
\url{https://doi.org/10.1109/TSM.2019.2938157}
\showDOI{\tempurl}


\bibitem[\protect\citeauthoryear{Rojas, Bastidas, and Cabrera}{Rojas
  et~al\mbox{.}}{2020}]%
        {Rojas2020}
\bibfield{author}{\bibinfo{person}{Elizabeth Rojas}, \bibinfo{person}{Viviana
  Bastidas}, {and} \bibinfo{person}{Christian Cabrera}.}
  \bibinfo{year}{2020}\natexlab{}.
\newblock \showarticletitle{{Cities-Board: A Framework to Automate the
  Development of Smart Cities Dashboards}}.
\newblock \bibinfo{journal}{\emph{IEEE Internet of Things Journal}}
  \bibinfo{volume}{7}, \bibinfo{number}{10} (\bibinfo{year}{2020}),
  \bibinfo{pages}{10128--10136}.
\newblock
\showISSN{23274662}
\urldef\tempurl%
\url{https://doi.org/10.1109/JIOT.2020.3002581}
\showDOI{\tempurl}


\bibitem[\protect\citeauthoryear{Romanovskaya}{Romanovskaya}{2019}]%
        {darkmode3}
\bibfield{author}{\bibinfo{person}{Viktoria Romanovskaya}.}
  \bibinfo{year}{2019}\natexlab{}.
\newblock \bibinfo{booktitle}{\emph{2019 UI/UX Design Trends You Should Know}}.
\newblock
\urldef\tempurl%
\url{https://medium.com/akveo-engineering/2019-ui-ux-design-trends-you-should-know-268b6bdbc0e3}
\showURL{%
Retrieved Jul 23, 2021 from \tempurl}


\bibitem[\protect\citeauthoryear{Runeson and H{\"{o}}st}{Runeson and
  H{\"{o}}st}{2009}]%
        {Runeson2009}
\bibfield{author}{\bibinfo{person}{Per Runeson} {and} \bibinfo{person}{Martin
  H{\"{o}}st}.} \bibinfo{year}{2009}\natexlab{}.
\newblock \showarticletitle{{Guidelines for conducting and reporting case study
  research in software engineering}}.
\newblock \bibinfo{journal}{\emph{Empirical Software Engineering}}
  \bibinfo{volume}{14}, \bibinfo{number}{2} (\bibinfo{year}{2009}),
  \bibinfo{pages}{131--164}.
\newblock
\showISSN{13823256}
\urldef\tempurl%
\url{https://doi.org/10.1007/s10664-008-9102-8}
\showDOI{\tempurl}


\bibitem[\protect\citeauthoryear{Sarikaya, Correll, Bartram, Tory, and
  Fisher}{Sarikaya et~al\mbox{.}}{2019}]%
        {Sarikaya2019}
\bibfield{author}{\bibinfo{person}{Alper Sarikaya}, \bibinfo{person}{Michael
  Correll}, \bibinfo{person}{Lyn Bartram}, \bibinfo{person}{Melanie Tory},
  {and} \bibinfo{person}{Danyel Fisher}.} \bibinfo{year}{2019}\natexlab{}.
\newblock \showarticletitle{{What do we talk about when we talk about
  dashboards?}}
\newblock \bibinfo{journal}{\emph{IEEE Transactions on Visualization and
  Computer Graphics}} \bibinfo{volume}{25}, \bibinfo{number}{1}
  (\bibinfo{year}{2019}), \bibinfo{pages}{682--692}.
\newblock
\showISSN{19410506}
\urldef\tempurl%
\url{https://doi.org/10.1109/TVCG.2018.2864903}
\showDOI{\tempurl}


\bibitem[\protect\citeauthoryear{Satyanarayan, Moritz, Wongsuphasawat, and
  Heer}{Satyanarayan et~al\mbox{.}}{2017}]%
        {Satyanarayan2017}
\bibfield{author}{\bibinfo{person}{Arvind Satyanarayan},
  \bibinfo{person}{Dominik Moritz}, \bibinfo{person}{Kanit Wongsuphasawat},
  {and} \bibinfo{person}{Jeffrey Heer}.} \bibinfo{year}{2017}\natexlab{}.
\newblock \showarticletitle{{Vega-Lite: A Grammar of Interactive Graphics}}.
\newblock \bibinfo{journal}{\emph{IEEE Transactions on Visualization and
  Computer Graphics}} \bibinfo{volume}{23}, \bibinfo{number}{1}
  (\bibinfo{year}{2017}), \bibinfo{pages}{341--350}.
\newblock
\showISSN{10772626}
\urldef\tempurl%
\url{https://doi.org/10.1109/TVCG.2016.2599030}
\showDOI{\tempurl}


\bibitem[\protect\citeauthoryear{Satyanarayan, Russell, Hoffswell, and
  Heer}{Satyanarayan et~al\mbox{.}}{2016}]%
        {Satyanarayan2016}
\bibfield{author}{\bibinfo{person}{Arvind Satyanarayan}, \bibinfo{person}{Ryan
  Russell}, \bibinfo{person}{Jane Hoffswell}, {and} \bibinfo{person}{Jeffrey
  Heer}.} \bibinfo{year}{2016}\natexlab{}.
\newblock \showarticletitle{{Reactive Vega: A Streaming Dataflow Architecture
  for Declarative Interactive Visualization}}.
\newblock \bibinfo{journal}{\emph{IEEE Transactions on Visualization and
  Computer Graphics}} \bibinfo{volume}{22}, \bibinfo{number}{1}
  (\bibinfo{year}{2016}), \bibinfo{pages}{659--668}.
\newblock
\showISSN{10772626}
\urldef\tempurl%
\url{https://doi.org/10.1109/TVCG.2015.2467091}
\showDOI{\tempurl}


\bibitem[\protect\citeauthoryear{Sendall and Kozaczynski}{Sendall and
  Kozaczynski}{2003}]%
        {Sendall2003}
\bibfield{author}{\bibinfo{person}{Shane Sendall} {and} \bibinfo{person}{Wojtek
  Kozaczynski}.} \bibinfo{year}{2003}\natexlab{}.
\newblock \showarticletitle{{Model transformation: The heart and soul of
  model-driven software development}}.
\newblock \bibinfo{journal}{\emph{IEEE Software}} \bibinfo{volume}{20},
  \bibinfo{number}{5} (\bibinfo{year}{2003}), \bibinfo{pages}{42--45}.
\newblock
\showISSN{07407459}
\urldef\tempurl%
\url{https://doi.org/10.1109/MS.2003.1231150}
\showDOI{\tempurl}


\bibitem[\protect\citeauthoryear{Sharma}{Sharma}{2017}]%
        {Mohit2017}
\bibfield{author}{\bibinfo{person}{Mohit~Kumar Sharma}.}
  \bibinfo{year}{2017}\natexlab{}.
\newblock \showarticletitle{A study of SDLC to develop well engineered
  software}.
\newblock \bibinfo{journal}{\emph{International Journal of Advanced Research in
  Computer Science}}  \bibinfo{volume}{8} (\bibinfo{year}{2017}).
\newblock
Issue 3.
\showISSN{0976-5697}
\urldef\tempurl%
\url{www.ijarcs.info}
\showURL{%
\tempurl}


\bibitem[\protect\citeauthoryear{Skorobogataya}{Skorobogataya}{2021}]%
        {darkmode2}
\bibfield{author}{\bibinfo{person}{Alexandra Skorobogataya}.}
  \bibinfo{year}{2021}\natexlab{}.
\newblock \bibinfo{booktitle}{\emph{Top 10 Dark-themed Admin Dashboards 2021
  You Need To See}}.
\newblock
\urldef\tempurl%
\url{https://www.akveo.com/blog/top-10-dark-themed-admin-dashboards-2021-you-need-to-see}
\showURL{%
Retrieved Jul 23, 2021 from \tempurl}


\bibitem[\protect\citeauthoryear{Topalian-Rivas, Wassermann, Severengiz, and
  Kr{\"{u}}ger}{Topalian-Rivas et~al\mbox{.}}{2020}]%
        {Topalian-Rivas2020}
\bibfield{author}{\bibinfo{person}{Garabet~A. Topalian-Rivas},
  \bibinfo{person}{Jonas Wassermann}, \bibinfo{person}{Mustafa Severengiz},
  {and} \bibinfo{person}{J{\"{o}}rg Kr{\"{u}}ger}.}
  \bibinfo{year}{2020}\natexlab{}.
\newblock \showarticletitle{{Automated dashboard generation for machine tools
  with OPC UA compatible sensors}}.
\newblock \bibinfo{journal}{\emph{IEEE International Conference on Emerging
  Technologies and Factory Automation, ETFA}}  \bibinfo{volume}{2020-Septe}
  (\bibinfo{year}{2020}), \bibinfo{pages}{1009--1012}.
\newblock
\showISBNx{9781728189567}
\showISSN{19460759}
\urldef\tempurl%
\url{https://doi.org/10.1109/ETFA46521.2020.9212136}
\showDOI{\tempurl}


\bibitem[\protect\citeauthoryear{Ullah and Babar}{Ullah and Babar}{2022}]%
        {Ullah2022}
\bibfield{author}{\bibinfo{person}{Faheem Ullah} {and} \bibinfo{person}{M.~Ali
  Babar}.} \bibinfo{year}{2022}\natexlab{}.
\newblock \showarticletitle{{On the scalability of Big Data Cyber Security
  Analytics systems}}.
\newblock \bibinfo{journal}{\emph{Journal of Network and Computer
  Applications}} \bibinfo{volume}{198}, \bibinfo{number}{November 2021}
  (\bibinfo{year}{2022}), \bibinfo{pages}{103294}.
\newblock
\showISSN{10958592}
\urldef\tempurl%
\url{https://doi.org/10.1016/j.jnca.2021.103294}
\showDOI{\tempurl}


\bibitem[\protect\citeauthoryear{V{\'{a}}zquez-Ingelmo, Garc{\'{i}}a-Holgado,
  Garc{\'{i}}a-Pe{\~{n}}alvo, and Ther{\'{o}}n}{V{\'{a}}zquez-Ingelmo
  et~al\mbox{.}}{2019}]%
        {Vazquez-Ingelmo2019}
\bibfield{author}{\bibinfo{person}{Andrea V{\'{a}}zquez-Ingelmo},
  \bibinfo{person}{Garc{\'{i}}a-Holgado},
  \bibinfo{person}{Garc{\'{i}}a-Pe{\~{n}}alvo}, {and}
  \bibinfo{person}{Ther{\'{o}}n}.} \bibinfo{year}{2019}\natexlab{}.
\newblock \showarticletitle{{Dashboard Meta-Model for Knowledge Management in
  Technological Ecosystem: A Case Study in Healthcare}}.
\newblock \bibinfo{journal}{\emph{Proceedings}} \bibinfo{volume}{31},
  \bibinfo{number}{1} (\bibinfo{year}{2019}), \bibinfo{pages}{44}.
\newblock
\showISSN{2504-3900}
\urldef\tempurl%
\url{https://doi.org/10.3390/proceedings2019031044}
\showDOI{\tempurl}


\bibitem[\protect\citeauthoryear{Vazquez-Ingelmo, Garcia-Penalvo, and
  Theron}{Vazquez-Ingelmo et~al\mbox{.}}{2019}]%
        {Vazquez-Ingelmo2019slr}
\bibfield{author}{\bibinfo{person}{Andrea Vazquez-Ingelmo},
  \bibinfo{person}{Francisco~J. Garcia-Penalvo}, {and} \bibinfo{person}{Roberto
  Theron}.} \bibinfo{year}{2019}\natexlab{}.
\newblock \showarticletitle{{Information Dashboards and Tailoring
  Capabilities-A Systematic Literature Review}}.
\newblock \bibinfo{journal}{\emph{IEEE Access}}  \bibinfo{volume}{7}
  (\bibinfo{year}{2019}), \bibinfo{pages}{109673--109688}.
\newblock
\showISSN{21693536}
\urldef\tempurl%
\url{https://doi.org/10.1109/ACCESS.2019.2933472}
\showDOI{\tempurl}


\bibitem[\protect\citeauthoryear{V{\'{a}}zquez-Ingelmo,
  Garc{\'{i}}a-Pe{\~{n}}alvo, Ther{\'{o}}n, {Amo Filv{\`{a}}}, and {Fonseca
  Escudero}}{V{\'{a}}zquez-Ingelmo et~al\mbox{.}}{2020a}]%
        {Vazquez-Ingelmo2020a}
\bibfield{author}{\bibinfo{person}{Andrea V{\'{a}}zquez-Ingelmo},
  \bibinfo{person}{Francisco~Jos{\'{e}} Garc{\'{i}}a-Pe{\~{n}}alvo},
  \bibinfo{person}{Roberto Ther{\'{o}}n}, \bibinfo{person}{Daniel {Amo
  Filv{\`{a}}}}, {and} \bibinfo{person}{David {Fonseca Escudero}}.}
  \bibinfo{year}{2020}\natexlab{a}.
\newblock \showarticletitle{{Connecting domain-specific features to source
  code: towards the automatization of dashboard generation}}.
\newblock \bibinfo{journal}{\emph{Cluster Computing}} \bibinfo{volume}{23},
  \bibinfo{number}{3} (\bibinfo{year}{2020}), \bibinfo{pages}{1803--1816}.
\newblock
\showISBNx{1058601903}
\showISSN{15737543}
\urldef\tempurl%
\url{https://doi.org/10.1007/s10586-019-03012-1}
\showDOI{\tempurl}


\bibitem[\protect\citeauthoryear{V{\'{a}}zquez-Ingelmo,
  Garc{\'{i}}a-Pe{\~{n}}alvo, Ther{\'{o}}n, and
  Garc{\'{i}}a-Holgado}{V{\'{a}}zquez-Ingelmo et~al\mbox{.}}{2020b}]%
        {Vazquez-Ingelmo2020}
\bibfield{author}{\bibinfo{person}{Andrea V{\'{a}}zquez-Ingelmo},
  \bibinfo{person}{Francisco~J. Garc{\'{i}}a-Pe{\~{n}}alvo},
  \bibinfo{person}{Roberto Ther{\'{o}}n}, {and} \bibinfo{person}{Alicia
  Garc{\'{i}}a-Holgado}.} \bibinfo{year}{2020}\natexlab{b}.
\newblock \showarticletitle{{Specifying information dashboards' interactive
  features through meta-model instantiation}}.
\newblock \bibinfo{journal}{\emph{CEUR Workshop Proceedings}}
  \bibinfo{volume}{2671} (\bibinfo{year}{2020}), \bibinfo{pages}{47--59}.
\newblock
\showISBNx{0000000158817}
\showISSN{16130073}


\bibitem[\protect\citeauthoryear{Vega}{Vega}{[n.\,d.]}]%
        {vegaandd3}
\bibfield{author}{\bibinfo{person}{Vega}.} \bibinfo{year}{[n.\,d.]}\natexlab{}.
\newblock \bibinfo{booktitle}{\emph{Vega and D3}}.
\newblock
\urldef\tempurl%
\url{https://vega.github.io/vega/about/vega-and-d3/}
\showURL{%
Retrieved Nov 1, 2021 from \tempurl}


\bibitem[\protect\citeauthoryear{Watson}{Watson}{2021}]%
        {darkmode}
\bibfield{author}{\bibinfo{person}{Kathyrn Watson}.}
  \bibinfo{year}{2021}\natexlab{}.
\newblock \bibinfo{booktitle}{\emph{Is Dark Mode Better for Your Eyes?}}
\newblock
\urldef\tempurl%
\url{https://www.healthline.com/health/is-dark-mode-better-for-your-eyes#takeaway}
\showURL{%
Retrieved Jul 23, 2021 from \tempurl}


\bibitem[\protect\citeauthoryear{{Wikipedia}}{{Wikipedia}}{2022}]%
        {wiki:json}
\bibfield{author}{\bibinfo{person}{{Wikipedia}}.}
  \bibinfo{year}{2022}\natexlab{}.
\newblock \bibinfo{title}{JSON}.
\newblock
\newblock
\urldef\tempurl%
\url{https://en.wikipedia.org/wiki/JSON}
\showURL{%
\tempurl}
\newblock
\shownote{[Online; accessed 22-April-2022]}.


\bibitem[\protect\citeauthoryear{Wohlin}{Wohlin}{2021}]%
        {Wohlin2021}
\bibfield{author}{\bibinfo{person}{Claes Wohlin}.}
  \bibinfo{year}{2021}\natexlab{}.
\newblock \showarticletitle{{Case Study Research in Software Engineering—It
  is a Case, and it is a Study, but is it a Case Study?}}
\newblock \bibinfo{journal}{\emph{Information and Software Technology}}
  \bibinfo{volume}{133}, \bibinfo{number}{January} (\bibinfo{year}{2021}),
  \bibinfo{pages}{10--12}.
\newblock
\showISSN{09505849}
\urldef\tempurl%
\url{https://doi.org/10.1016/j.infsof.2021.106514}
\showDOI{\tempurl}


\bibitem[\protect\citeauthoryear{Yin}{Yin}{2003}]%
        {Yinbook2003}
\bibfield{author}{\bibinfo{person}{Robert~K. Yin}.}
  \bibinfo{year}{2003}\natexlab{}.
\newblock \showarticletitle{Case Study Research Design and Methods: Applied
  Social Research and Methods Series}. In \bibinfo{booktitle}{\emph{Case Study
  Research Design and Methods: Applied Social Research and Methods Series}}.
  \bibinfo{publisher}{Sage Publications}.
\newblock


\bibitem[\protect\citeauthoryear{Ünal Aksu, del Río-Ortega, Resinas, and
  Reijers}{Ünal Aksu et~al\mbox{.}}{2019}]%
        {Aksu2019}
\bibfield{author}{\bibinfo{person}{Ünal Aksu}, \bibinfo{person}{Adela del
  Río-Ortega}, \bibinfo{person}{Manuel Resinas}, {and}
  \bibinfo{person}{Hajo~A. Reijers}.} \bibinfo{year}{2019}\natexlab{}.
\newblock \showarticletitle{An approach for the automated generation of
  engaging dashboards}.
\newblock \bibinfo{journal}{\emph{Lecture Notes in Computer Science (including
  subseries Lecture Notes in Artificial Intelligence and Lecture Notes in
  Bioinformatics)}}  \bibinfo{volume}{11877 LNCS}, \bibinfo{pages}{363--384}.
\newblock
\showISBNx{9783030332457}
\showISSN{16113349}
\urldef\tempurl%
\url{https://doi.org/10.1007/978-3-030-33246-4_24}
\showDOI{\tempurl}


\end{thebibliography}

\newpage


\section*{Appendix}

\begin{table}[h] \caption{\label{tab:inclusionExclusion}Inclusion (I) and exclusion (E) criteria that applied when selecting dashboards.}
{%
\begin{tabular}{l|p{12cm}}
\hline
& \textbf{Inclusion criteria}\\

 \hline
I1 &	The dashboard should visualize and monitor critical information and conditions.\\
 \hline
I2	& The dashboard should help extract, explain, understand knowledge, and enhancing decision-making ability.\\
 \hline
I3 & The dashboard should use multiple data representation formats, and have interactive features.\\
\hline
& \textbf{Exclusion criteria}\\
\hline
E1 &	The dashboard over-designed or over-coded for particular purpose is excluded.\\ \hline
E2 &	The dashboard designed for mobile-only is excluded.\\\hline
E3 &	Duplicated dashboard design or similar design from the same source is excluded.\\\hline
E4 &	Sectional or uncompleted dashboard is excluded.\\
\hline
\end{tabular}%
} 
\end{table}

\begin{figure}[h]
\begin{minipage}[h]{0.8\textwidth}
\centering
\includegraphics[width=\textwidth]{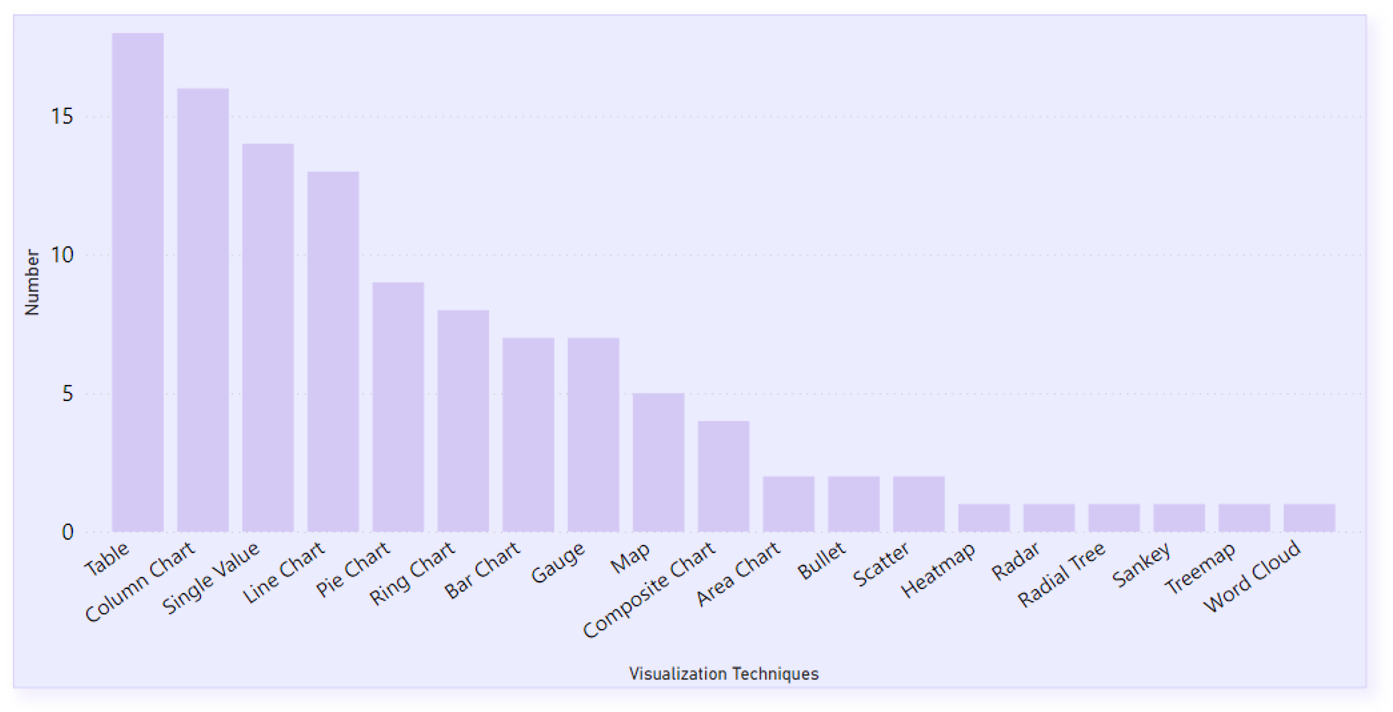}
\caption{Distribution of visualization techniques used in the real-world 31 cyber security dashboards.}
\label{vistech}
\end{minipage}\hfill%
\end{figure}


\newpage

\begin{table}[h] \caption{\label{tab:dataextration}Statistics and diversity of the case.}
\begin{tabular}{cccc}
\hline
\textbf{Item}                                   & \textbf{Category} & \textbf{Number} & \textbf{Percent} \\ \hline
\multirow{2}{*}{Purpose}                        & Monitoring        & 28           & 90.32            \\ 
                                                & Reporting         & 7            & 22.58            \\ \hline
\multirow{3}{*}{Stakeholder}                    & Analysts          & 24           & 77.41            \\ 
                                                & Executives        & 8            & 25               \\ 
                                                & Board Members     & 2            & 6.45             \\ \hline
\multirow{3}{*}{Time Sensitivity}               & High              & 5            & 16.13            \\ 
                                                & Medium            & 10           & 32.16            \\ 
                                                & Low               & 16           & 51.61            \\ \hline
\multirow{3}{*}{Decision Support Dimension}     & Strategic         & 11           & 35.48            \\ 
                                                & Tactical          & 18           & 58.06            \\ 
                                                & Operational       & 2            & 6.45             \\ \hline
\multirow{3}{*}{Situational Awareness Level}    & Perception        & 22           & 70.97            \\ 
                                                & Comprehension     & 9            & 29.03            \\ 
                                                & Projection        & 0            & 0                \\ \hline
\multirow{4}{*}{Layout}                         & Standard Grid \includegraphics[height=2ex]{image/l1.png}     & 3            & 9.68             \\ 
                                                & Row Oriented \includegraphics[height=2ex]{image/l3.png}      & 18           & 58.06            \\ 
                                                & Column Oriented \includegraphics[height=2ex]{image/l4.png}   & 4            & 12.9             \\ 
                                                & Disordered \includegraphics[height=2ex]{image/l2.png}        & 6            & 19.35            \\ \hline
\multirow{2}{*}{Theme}                          & Light             & 22           & 70.97            \\ 
                                                & Dark              & 9            & 29.03            \\ \hline

\multirow{2}{*}{Interaction}                    & Dashboard Level   & 19           & 61.29            \\ 
                                                & Widget Level      & 16           & 51.61            \\ \hline
\multirow{3}{*}{Widget Number}                  & Max               & 18           &                  \\ 
                                                & Average           & 8            &                  \\ 
                                                & Min               & 3            &                  \\ \hline
\multirow{3}{*}{Visualization Technique Number} & Max               & 7            &                  \\  
                                                & Average           & 4            &                  \\ 
                                                & Min               & 2            &                  \\ \hline
\end{tabular}
\end{table}

\newpage

\received{February 2022}
\received[accepted]{April 2022}

\begin{figure}[t]
\begin{minipage}[t]{0.9\textwidth}
\centering
\includegraphics[width=1\textwidth]{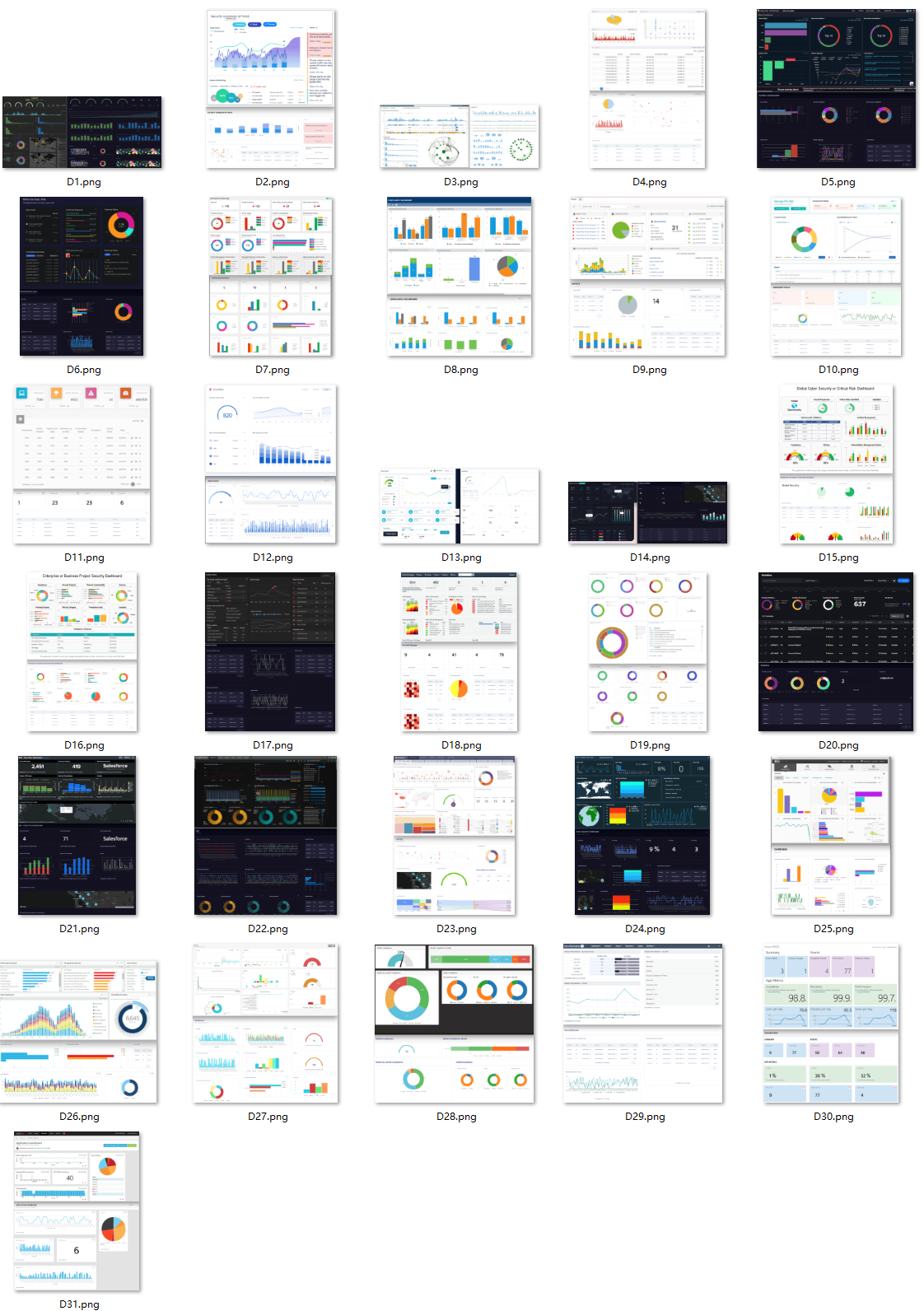}
\caption{All of the 31 real-world cyber security dashboards have been reconstructed by Mod2Dash.}
\label{alldashboards}
\end{minipage}\hfill%
\end{figure}

\end{document}